\documentclass[draft, table]{agujournal2019}
\usepackage{url} %this package should fix any errors with URLs in refs.
\usepackage{lineno}
\usepackage{amsmath} %For cases
\usepackage{amssymb}
\usepackage{amsfonts}
\usepackage[inline]{trackchanges} %for better track changes. finalnew option will compile document with changes incorporated.
\usepackage{soul}
\usepackage{multirow}
\usepackage{xcolor} % For zebra lines
\usepackage[inline]{trackchanges} %for better track changes. finalnew option will compile document with changes incorporated.
\usepackage{gensymb}
\usepackage{textcomp, array, multirow, bigdelim, makecell, rotating} % Curly bracket extending multiple rows
\usepackage{booktabs} % For nice looking tables
\usepackage{xr} % To be able to refer to figures and tables in Supplemental_Information.tex
\usepackage{bm}
\usepackage[shortlabels]{enumitem}

\draftfalse

\journalname{Journal of Advances in Modeling Earth Systems (JAMES)}

\begin{document}

%% ------------------------------------------------------------------------ %%
%  Title
%
% (A title should be specific, informative, and brief. Use
% abbreviations only if they are defined in the abstract. Titles that
% start with general keywords then specific terms are optimized in
% searches)
%
%% ------------------------------------------------------------------------ %%

% Example: \title{This is a test title}

%\graphicspath{{figures/}}

\title{Data-Driven Equation Discovery of a Cloud Cover Parameterization}

%% ------------------------------------------------------------------------ %%
%
%  AUTHORS AND AFFILIATIONS
%
%% ------------------------------------------------------------------------ %%

% Authors are individuals who have significantly contributed to the
% research and preparation of the article. Group authors are allowed, if
% each author in the group is separately identified in an appendix.)

% List authors by first name or initial followed by last name and
% separated by commas. Use \affil{} to number affiliations, and
% \thanks{} for author notes.
% Additional author notes should be indicated with \thanks{} (for
% example, for current addresses).

% Example: \authors{A. B. Author\affil{1}\thanks{Current address, Antartica}, B. C. Author\affil{2,3}, and D. E.
% Author\affil{3,4}\thanks{Also funded by Monsanto.}}

\authors{Arthur Grundner\affil{1,2}, Tom Beucler\affil{3}, Pierre Gentine\affil{2}, and Veronika Eyring\affil{1,4}}

% \affiliation{1}{First Affiliation}
% \affiliation{2}{Second Affiliation}
% \affiliation{3}{Third Affiliation}
% \affiliation{4}{Fourth Affiliation}

\affiliation{1}{Deutsches Zentrum für Luft- und Raumfahrt e.V. (DLR), Institut für Physik der Atmosphäre, Oberpfaffenhofen, Germany}
\affiliation{2}{Center for Learning the Earth with Artificial Intelligence And Physics (LEAP), Columbia University, New York, NY, USA}
\affiliation{3}{Institute of Earth Surface Dynamics, University of Lausanne, Lausanne, Switzerland}
\affiliation{4}{University of Bremen, Institute of Environmental Physics (IUP), Bremen, Germany}
%(repeat as many times as is necessary)

%% Corresponding Author:
% Corresponding author mailing address and e-mail address:

% (include name and email addresses of the corresponding author.  More
% than one corresponding author is allowed in this LaTeX file and for
% publication; but only one corresponding author is allowed in our
% editorial system.)

% Example: \correspondingauthor{First and Last Name}{email@address.edu}

\correspondingauthor{Arthur Grundner}{arthur.grundner@dlr.de}

%% Keypoints, final entry on title page.

%  List up to three key points (at least one is required)
%  Key Points summarize the main points and conclusions of the article
%  Each must be 140 characters or fewer with no special characters or punctuation and must be complete sentences

% Example:
% \begin{keypoints}
% \item	List up to three key points (at least one is required)
% \item	Key Points summarize the main points and conclusions of the article
% \item	Each must be 140 characters or fewer with no special characters or punctuation and must be complete sentences
% \end{keypoints}

\begin{keypoints} % Checked by PG, TB. Not more than 140 chars per KP.
\item We systematically derive and evaluate cloud cover parameterizations of various complexity from global storm-resolving simulation output % 135 chars
\item Using symbolic regression combined with physical constraints, we find a new interpretable equation balancing performance and simplicity % 135 chars
% Alternatively (if one wants to avoid featuring): Using symbolic regression, we find a new analytical cloud cover scheme representing an excellent trade-off between performance and simplicity (139 chars)
\item Our data-driven cloud cover equation can be retuned with few samples, facilitating transfer learning to generalize to other realistic data % 138 chars

% Using symbolic regression with physical constraints, we find a new interpretable equation balancing performance and simplicity

% \item \textit{These are the best models on DYAMOND:}
% \item \textit{These are the best model when transferred to ERA5:}
% \item \textit{Our best model works well across regimes and data sets because XYZ (physical analysis section, which should include physical interpretation of the trainable coefficients)}
% \item \textit{Simple analytical models can be re-tuned with less data, facilitating transfer learning (Problem is that it mostly requires retraining for decent performance, interpretability/transparency component, also a lot less retraining required, idea of fixed functional structure)}
% \item \textit{Regime-based evaluation - tied to the scientific point we want to make about regimes (decision tree) - so far mostly helps for evaluation (clearly shows the limit of simple analytical equations when it comes to fiting the PDFs)}
% \item \textit{Main scientific findings (could be derived from Pareto frontier, could be about regimes but also nonlinearity, feature selection, etc.)}
\end{keypoints}

\begin{abstract} % 250 word limit!
A promising method for improving the representation of clouds in climate models, and hence climate projections, is to develop machine learning-based parameterizations using output from global storm-resolving models. While neural networks can achieve state-of-the-art performance within their training distribution, they can make unreliable predictions outside of it. Additionally, they often require post-hoc tools for interpretation. To avoid these limitations, we combine symbolic regression, sequential feature selection, and physical constraints in a hierarchical modeling framework. This framework allows us to discover new equations diagnosing cloud cover from coarse-grained variables of global storm-resolving model simulations. These analytical equations are interpretable by construction and easily transferable to other grids or climate models. Our best equation balances performance and complexity, achieving a performance comparable to that of neural networks ($R^2=0.94$) while remaining simple (with only 11 trainable parameters). It reproduces cloud cover distributions more accurately than the Xu-Randall scheme across all cloud regimes (Hellinger distances $<0.09$), and matches neural networks in condensate-rich regimes. When applied and fine-tuned to the ERA5 reanalysis, the equation exhibits superior transferability to new data compared to all other optimal cloud cover schemes. Our findings demonstrate the effectiveness of symbolic regression in discovering interpretable, physically-consistent, and nonlinear equations to parameterize cloud cover. % 198 words

% Informed by a sequential feature selection approach, the equation depends on relative humidity, temperature, cloud ice/water, and the vertical derivative of relative humidity. -- VE?

\end{abstract}

\section*{Plain Language Summary} % Max. 200 words?
In climate models, cloud cover is usually expressed as a function of coarse, pixelated variables. Traditionally, this functional relationship is derived from physical assumptions. In contrast, machine learning approaches, such as neural networks, sacrifice interpretability for performance. In our approach, we use high-resolution climate model output to learn a hierarchy of cloud cover schemes from data. To bridge the gap between simple statistical methods and machine learning algorithms, we employ a symbolic regression method. Unlike classical regression, which requires providing a set of basis functions from which the equation is composed of, symbolic regression only requires mathematical operators (such as $+$, $\times$) that it learns to combine. By using a genetic algorithm, inspired by the process of natural selection, we discover an interpretable, nonlinear equation for cloud cover. This equation is simple, performs well, satisfies physical principles, and outperforms other algorithms when applied to new observationally-informed data. % 168 words

\section{Introduction}
% What is the underlying general problem? How is it commonly solved?
Due to computational constraints, climate models used to make future projections spanning multiple decades typically have horizontal resolutions of 50\textendash{}100\,km \cite{eyring2021}. The coarse resolution necessitates the parameterization of many subgrid-scale processes (e.g., radiation, microphysics), which have a significant effect on model forecasts \cite{stensrud2009}. Climate models, such as the state-of-the-art ICOsahedral Non-hydrostatic (ICON) model, exhibit long-standing systematic biases, especially related to cloud parameterizations \cite{crueger2018icon, giorgetta2018}. A fundamental component of the cloud parameterization package in ICON is its cloud cover scheme, which, in its current form, diagnoses fractional cloud cover from large-scale variables in every grid cell \cite{giorgetta2018, mauritsen2019}. As cloud cover is directly used in the radiation \cite{pincus2013} and cloud microphysics \cite{lohmann1996} parameterizations of ICON, its estimate directly influences the energy balance and the statistics of water vapor, cloud ice, and cloud water. The current cloud cover scheme in ICON, based on \citeA{sundqvist1989}, nevertheless makes some crude empirical assumptions, such as a near-exclusive emphasis on relative humidity (see \citeA{grundner2022} for further discussion). These assumptions may impede the search for a parameterization that faithfully captures the available data.

% ML Introduction
With the extended availability of high-fidelity data and increasingly sophisticated machine learning (ML) methods, ML algorithms have been developed for the parameterization of clouds and convection (e.g., \citeA{brenowitz2018, gentine2018, krasnopolsky2013, ogorman2018}; see reviews by \citeA{beucler2022machine} and \citeA{gentine2021deep}). High-resolution atmospheric simulations on storm-resolving scales (horizontal resolutions of a few kilometers) resolve deep convective processes explicitly \cite{weisman1997resolution}, and provide useful training data with an improved physical representation of clouds and convection \cite{hohenegger2020, stevens2020added}. There are only few approaches that learn parameterizations directly from observations (e.g., \citeA{mccandless2022machine}), as these are challenged by the sparsity and noise of observations \cite{rasp2018, trenberth2009earth}. Therefore, a two-step process might be required, in which the statistical model structure is first learned on high-resolution modeled data before its parameters are fine-tuned on observations (transfer learning), leveraging the advantage of the consistency of the modeled data for the initial training stage before having to deal with noisier observational data.

% maybe split up!
Neural networks and random forests have been routinely used for ML-based parameterizations. Unlike traditional regression approaches, they are not limited to a particular functional form provided by combining a set of basis functions. They are usually fast at inference time and can be trained with very little domain knowledge. However, this versatility comes at the cost of interpretability as explainable artificial intelligence (XAI) methods still face major challenges \cite{kumar2020problems, molnar2021interpretable}. Given this limitation, we ask:
% TB: I thought the transition to the paragraph below was a bit abrupt so I added an overall, guiding question, which could probably be improved upon
Can we create data-driven cloud cover schemes that are interpretable by construction without renouncing the high data fidelity of neural networks?

% Introduction of our own approach
Here, we use a hierarchical modeling approach to systematically derive and evaluate a family of cloud cover (interpreted as the cloud area fraction) schemes, ranging from traditional physical (but semi-empirical) schemes and simple regression models to neural networks. We evaluate them according to their Pareto optimality (i.e., whether they are the best performing model for their complexity). To bridge the gap between simple equations and high-performance neural networks, we apply equation discovery in a data-driven manner using state-of-the-art symbolic regression methods. In symbolic regression, as opposed to regular regression, the user first specifies a set of mathematical operators instead of a set of basis functions. For instance, including division as a mathematical operator may introduce rational nonlinearities, whose ubiquity and importance have been illustrated, e.g., in \citeA{kaheman2020sindy}. Based on these operators, the symbolic regression library creates a random initial population of equations \cite{schmidt2009distilling}. Inspired by the process of natural selection in the theory of evolution, symbolic regression is usually implemented as a genetic algorithm that iteratively applies genetically motivated operations (selection, crossover, mutation) to the set of candidate equations. At each step, the equations are ranked based on their performance and simplicity, so that the top equations can be selected to be included in the next population \cite{smits2005pareto}. Advantages of training/discovering analytical models instead of neural networks include an immediate view of model content (e.g., whether physical constraints are satisfied) and the ability to analyze the model structure directly using powerful mathematical tools (e.g., perturbation theory, numerical stability analysis). Additionally, analytical models are straightforward to communicate to the broader scientific community, to implement numerically, and fast to execute given the existence of optimized implementations of well-known functions.

% Reviewing data-driven equation discovery
% I would keep it relatively short (I'm already discussing SRBench, PySR, GPGOMEA, DSR, AIFeynman in Sec~3.3)
% I'm also picking up Zanna, Ross again in the conclusion
To our knowledge, \citeA{zanna2020data} marks the first usage of automated, data-driven equation discovery for climate applications. Training on highly idealized data, they used a sparse regression technique called relevance vector machine to find an analytical model that parameterizes ocean eddies. In sparse regression, the user defines a library of terms, and the algorithm determines a linear combination of those terms that best matches the data while including as few terms as possible \cite{brunton2016discovering, rudy2017data, zhang2018robust, champion2019data}. In a follow-up paper, \citeA{ross2023benchmarking} employed symbolic regression to discover an improved equation, again trained on idealized data, that performs similarly well as neural networks across various metrics and has greater generalization capability. Nonetheless, they had to assume that the equation was linear in terms of its free/trainable parameters and additively separable as their method included an iterative approach to select suitable terms. For the selection of terms, they took a human-in-the-loop approach rather than solely relying on the genetic algorithm. Additionally, the final discovered equation relied on high-order spatial derivatives, which may not be feasible to compute in a climate model. To prevent this issue, we only permit features we can either access or easily derive in the climate model.

% \textcolor{blue}{TB: Could review data-driven equation discovery for atmospheric and oceanic sciences, its challenges, and contrast your approach to the recent paper from A. Ross and Laure. There's the question of whether we would like to review equation discovery more broadly. That would include the Brunton/Champion papers + \cite{udrescu2020ai,long2019pde,schmidt2009distilling,kim2020integration}, \cite{rackauckas2020universal} for universal differential equations, which includes symbolic regression of the forcing of ODEs/PDEs, symbolic regression libraries \cite{cranmer2020pysr,kim2021interactive}, various equation learning frameworks \cite{zhang2018robust,cranmer2020discovering,sahoo2018learning,jin2019bayesian}. We can also already review Pareto-optimal models \cite{smits2005pareto,udrescu2020ai} in the introduction, although this could be directly reviewed in section 5.1. Paper by Bolton and Zanna, DARPA?, SRBench, ...}

% Key Questions 
Guiding questions for this study include: Using symbolic regression, can we automatically discover a physically consistent equation for cloud cover whose performance is competitive with that of neural networks? Given that modern symbolic regression libraries can handle higher computational overhead, we want to relax prior assumptions of linearity or separability of the equation. Then, what can we learn about the cloud cover parameterization problem by sequentially selecting performance-maximizing features in different predictive models? Finally, how much better do simple models generalize and/or transfer to more realistic data sets? 

% Outline
We first introduce the data sets used for training, validation and testing (Sec~\ref{sec:data}), the diverse data-driven models used in this study (Sec~\ref{sec:model_families}), and evaluation metrics (Sec~\ref{sec:model_eval}), before studying the feature rankings, performances and complexities of the different models (Sec~\ref{sec:results_dyamond}). We investigate their ability to reproduce cloud cover distributions (Sec~\ref{sec:cloud_regimes}), transfer to higher resolutions (Sec~\ref{sec:transfer_res}), and adapt to the ERA5 reanalysis (Sec~\ref{sec:transfer}). We conclude with an analysis of the best analytical model we found using symbolic regression (Sec~\ref{sec:pysr_phys_int}).

%% General structure of an introduction

% \textsc{What is the underlying general problem? [Situation]} (Three sentences) (\textit{What is the problem you are going to tackle?}, \textit{What research area is addressed?}) 

% \textsc{How is it commonly solved? [Situation]}

% \textsc{Problems with the common solution [Complication]}

% \textsc{What is our approach, with a short explanation and what are some key advantages? [RQ]} (\textit{How did you solve the problem?}, \textit{What methods/solutions are applied?})

% \textsc{Short overview of what we do, with reference to previous studies [RQ]} (\textit{What are the research voids?}, \textit{What is the research question?}, \textit{How is the research void addressed?})

% \textsc{What key questions are we going to tackle? Summarize what we will write about [Solution]} (\textit{A final paragraph should describe how the paper will proceed})

\section{Data \label{sec:data}}
In this section, we introduce the two data sets used to train and benchmark our cloud cover schemes: We first use storm-resolving ICON simulations to train high-fidelity models (Sec~\ref{sec:data_dyamond}), before testing these models' transferability to the ERA5 meteorological reanalysis, which is more directly informed by observations (Sec~\ref{sub:ERA5}).

\subsection{Global Storm-Resolving Model Simulations (DYAMOND) \label{sec:data_dyamond}}

% Xu-Randall, Texeira, Sundqvist: SR data (discard spin-up, downsample) -> Train/Valid split \\
% Linear fits: SR data -> normalized acc. to training set -> Train/Valid split \\
% Polynomials: SR data -> normalized acc. to training set -> Train/Valid split \\
% SFS NNs: SR data -> normalized acc. to training set -> Train/Valid split \\
% PySR equations: SR data -> normalized acc. to training set -> Train/Valid split \\
% NNs: Keep spin-up, undersampling differs. Train/Valid sets completely different than for all other models!

% https://gitlab.dkrz.de/icon-ml/ml-based-parameterizations-for-icon/-/issues/261
As the source for our training data, we use output from global storm-resolving ICON simulations performed as part of the DYnamics of the Atmospheric general circulation Modeled On Non-hydrostatic Domains (DYAMOND) project. The project's first phase (`DYAMOND Summer') included a simulation starting from August 1, 2016 \cite{stevens2019}, while the second phase (`DYAMOND Winter') was initialized on January 20, 2020 \cite{duras2021dyamond}. In both phases, the ICON model simulated 40 days, providing three-hourly output on a grid with a horizontal resolution of $2.47\,$km.

Following the methodology of \citeA{grundner2022}, we coarse-grain the DYAMOND data to an ICON grid with a typical climate model horizontal grid resolution of $\approx\,80\,$km. Vertically, we coarse-grain the data from 58 to 27 layers below an altitude of $21\,$km, which is the maximum altitude with clouds in the data set. For cloud cover, we first estimate the vertically maximal cloud cover values in each low-resolution grid cell before horizontally coarse-graining the resulting field. For all other variables, we take a three-dimensional integral over the high-resolution grid cells overlapping a given low-resolution grid cell. For details, we refer the reader to Appendix A of \citeA{grundner2022}. Due to the sequential processing of some parameterization schemes in the ICON model, condensate-free clouds can occur in the simulation output. To instead ensure consistency between cloud cover and the other model variables, we follow \citeA{giorgetta2022} and manually set the cloud cover in the high-resolution grid cells to $100\%$ when the cloud condensate mixing ratio exceeds $10^{-6}\,$kg/kg and to $0\%$ otherwise.

We remove the first ten days of `DYAMOND Summer' and `DYAMOND Winter' as spin-up, and discard columns that contain NaNs ($3.15\%$ of all columns). From the remainder, we keep a random subset of $28.5\%$ of the data, while removing predominantly cloud-free cells to mitigate a class imbalance in the output (`undersampling' step). We then split the data into a training and a validation set, the latter of which is used for early stopping. To avoid high correlations between the training and validation sets, we divide the data set into six temporally connected parts. We choose the union of the second ($\approx\,$Aug 21\textendash{}Sept 1, 2016) and the fifth ($\approx\,$Feb 9\textendash{}Feb 19, 2020) part to create our validation set. For all models except the traditional schemes, we additionally normalize models' features (or `inputs') so that they have zero mean and unit variance on the training set.

We define a set of 24 features $\mathcal{F}$ that the models (discussed in Sec~\ref{sec:model_families}) can choose from. For clarity, we decompose $\mathcal{F}$ into three subsets: $\mathcal{F} \overset{\mathrm{def}}{=} \mathcal{F}_1 \cup \mathcal{F}_2 \cup \mathcal{F}_3$. The first subset, $\mathcal{F}_1 \overset{\mathrm{def}}{=} \{U, q_v, q_c, q_i, T, p, \mathrm{RH}\}$ groups the horizontal wind speed $U [m/s]$ and thermodynamic variables known to influence cloud cover, namely specific humidity $q_v\, [kg/kg]$, cloud water and ice mixing ratios $q_c\, [kg/kg]$ and $q_i\, [kg/kg]$, temperature $T\, [K]$, pressure $p\, [Pa]$ , and relative humidity $\mathrm{RH}$ with respect to water, approximated as:
\begin{equation}
\label{eq:rh}
    \mathrm{RH} \approx 0.00263 \frac{p}{1\mathrm{Pa}} q_v \exp \left[ \frac{17.67(273.15\mathrm{K} - T)}{T-29.65\mathrm{K}} \right].
\end{equation}
The second subset $\mathcal{F}_2$ contains the first and second vertical derivatives of all features in $\mathcal{F}_1$. These derivatives are computed by fitting splines to every vertical profile of a given variable and differentiating the spline at the grid level heights to obtain derivatives on the irregular vertical grid. Finally, the third subset $\mathcal{F}_3 \overset{\mathrm{def}}{=} \{z, \text{land}, p_s\}$ includes geometric height $z\, [m]$ and the only two-dimensional variables, i.e., land fraction and surface pressure $p_s\, [Pa]$.

In \citeA{grundner2022} we found it sufficient to diagnose cloud cover using information from the close vertical neighborhood of a grid cell. By utilizing vertical derivatives to incorporate this information, we ensure the applicability of our cloud cover schemes to any vertical grid. Since our feature set $\mathcal{F}$ contains all features appearing in our three baseline `traditional' parameterizations (see Sec~\ref{sec:ex_schemes}), we deem it comprehensive enough for the scope of our study. 

\subsection{Meteorological Reanalysis (ERA5) \label{sub:ERA5}}

To test the transferability of our cloud cover schemes to observational data, we also use the ERA5 meteorological reanalysis \cite{hersbach2018}. We sample the first day of each quarter in 1979\textendash{}2021 at a three-hourly resolution. The days from 2000\textendash{}2006 are taken from ERA5.1, which uses an improved representation of the global-mean temperatures in the upper troposphere and stratosphere. Depending on the ERA5 variable, they are either stored on an N320 reduced Gaussian (e.g., for cloud cover) or a T639 spectral (e.g., for temperature) grid. Using the CDO package \cite{schulzweida_uwe_2019_3539275}, we first remap all relevant variables to a regular Gaussian grid, and then to the unstructured ICON grid described in Sec~\ref{sec:data_dyamond}. Vertically, we coarse-grain from approximately 90 to 27 layers.

\begin{figure}
\centering
\hspace*{-4cm}\includegraphics[width=1.5\textwidth]{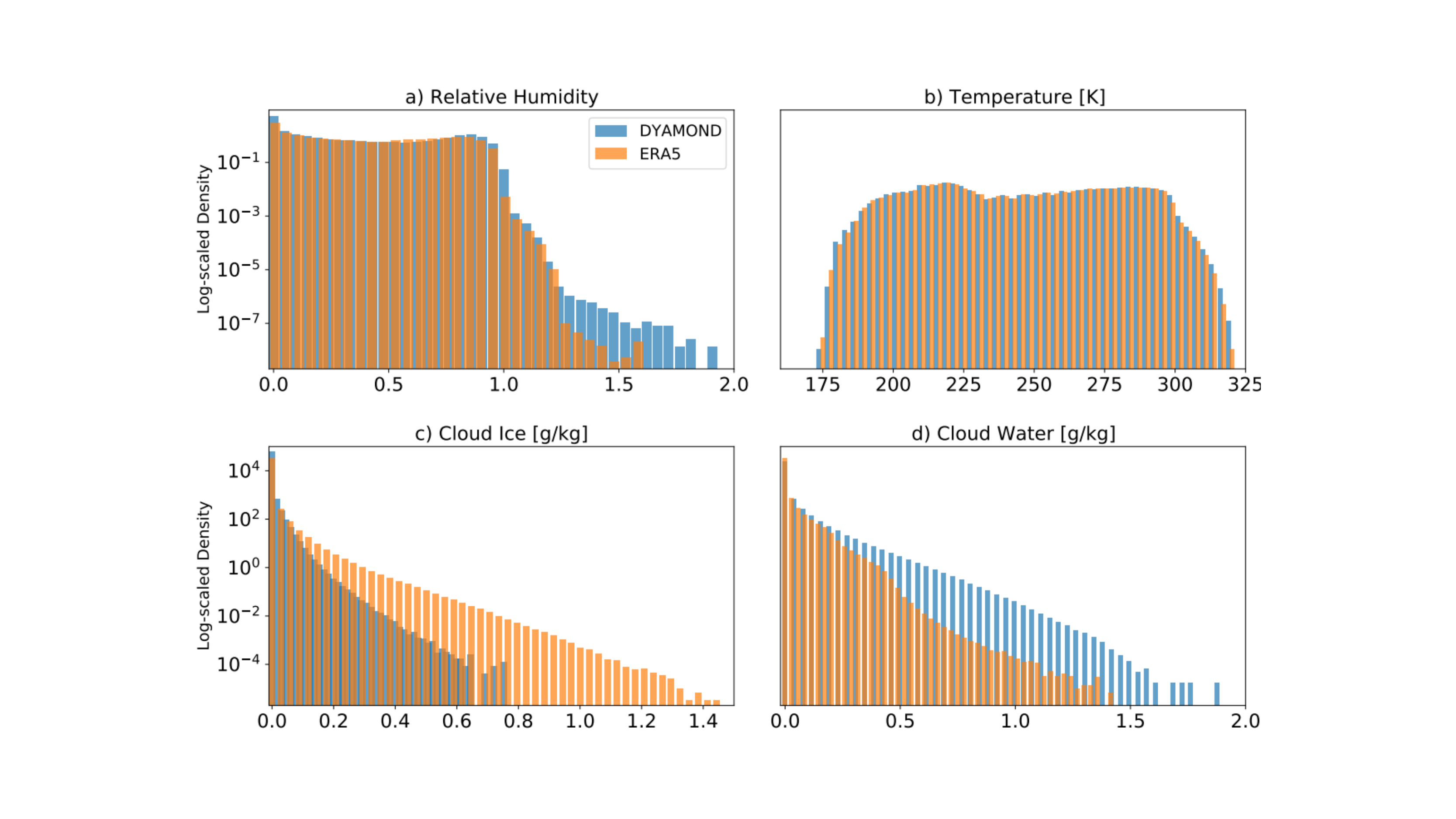}
\caption{A comparison of the univariate distributions of four variables from the coarse-grained DYAMOND and ERA5 data sets. The y-axes are scaled logarithmically to visualize the distributions' tails. While cloud ice is often larger in our processed ERA5 data set, cloud water tends to be smaller than in the DYAMOND data. The distributions of temperature and relative humidity are comparable.}
\label{fig:era5}
\end{figure}

The univariate distributions of important features such as cloud water and ice do not match between the (coarse-grained) DYAMOND and (processed) ERA5 data. The maximal cloud ice values that are attained in the ERA5 data set are twice as large as in the DYAMOND data. We illustrate this in Fig~\ref{fig:era5}, next to a comparison of the distributions of cloud water, relative humidity and temperature. Due to differences in the distributions of cloud ice, cloud water and relative humidity, we consider our processed ERA5 data a challenging data set to generalize to.

\section{Data-Driven Modeling}
\label{sec:model_families}
We now introduce a family of data-driven cloud cover schemes. We adopt a hierarchical modeling approach and start with models that are interpretable by construction, i.e., linear models, polynomials, and traditional schemes. As a second step, we mostly focus on performance and therefore train deep neural networks (NNs) on the DYAMOND data. To bridge the gap between the best-performing and most interpretable models, we use symbolic regression to discover analytical cloud cover schemes from data. These schemes are complex enough to include relevant nonlinearities while remaining interpretable.

\subsection{Existing Schemes}
\label{sec:ex_schemes}
We first introduce three traditional diagnostic schemes for cloud cover and train them using the BFGS \cite{nocedal1999numerical} and Nelder-Mead \cite{gao2012implementing} unconstrained optimizers (which outperform grid search methods in our case), each time choosing the model that minimizes the mean squared error (MSE) on the validation set. Before doing so, we multiply the output of each of the three schemes by $100$ to obtain percent cloud cover values. The first is the Sundqvist scheme \cite{sundqvist1989}, which is currently implemented in the ICON climate model \cite{giorgetta2018}. The Sundqvist scheme expresses cloud cover as a monotonically increasing function of relative humidity. It assumes that cloud cover can only exist if relative humidity exceeds a critical relative humidity threshold $\mathrm{RH}_0$, which itself is a function of the fraction between surface pressure and pressure: If
\begin{equation}
\mathrm{RH} > \mathrm{RH}_0 \stackrel{\mathrm{def}}{=} \mathrm{RH}_{\mathrm{0, top}} + (\mathrm{RH}_{\mathrm{0, surf}} - \mathrm{RH}_{\mathrm{0, top}})\exp(1-\left(p_s/p\right)^n),
\end{equation}
then the Sundqvist cloud cover is given by
\begin{equation}
{\cal C}_{\mathrm{Sundqvist}}\overset{\mathrm{def}}{=}1-\sqrt{\frac{\min\{\mathrm{RH},\mathrm{RH}_{\mathrm{sat}}\}-\mathrm{RH_{sat}}}{\mathrm{RH}_{0}-\mathrm{RH_{sat}}}}
.
\end{equation}
The Sundqvist scheme has four tunable parameters $\left\{ \mathrm{RH}_{\mathrm{0, surf}},\mathrm{RH}_{\mathrm{0, top}},\mathrm{RH}_{\mathrm{sat}},n\right\} 
$. As properly representing marine stratocumulus clouds in the Sundqvist scheme might require a different treatment (see, e.g., \citeA{mauritsen2019}), we allow these parameters to differ between land and sea, which we separate using a land fraction threshold of 0.5. 

The second scheme is a simplified version of the Xu-Randall scheme \cite{xu1996semiempirical}, which was found to outperform the Sundqvist scheme on CloudSat data \cite{wang2023evaluating}. It additionally depends on cloud water and ice, ensuring that cloud cover is 0 in condensate-free grid cells. It can be formulated as
\begin{equation}
\label{eq:xu-randall}
{\cal C}_\mathrm{Xu-Randall} \stackrel{\mathrm{def}}{=} \min\{\mathrm{RH}^{\beta}(1-\exp(-\alpha(q_c + q_i))), 1\}.
\end{equation}
The Xu-Randall scheme has only two tuning parameters: $\{\alpha, \beta\}$.

The third scheme was introduced in \citeA{teixeira2001cloud} for subtropical boundary layer clouds. Teixeira arrived at a diagnostic relationship for cloud cover by equating a cloud production term from detrainment and a cloud erosion term from turbulent mixing with the environment. We can express the Teixeira scheme as
\begin{equation}
{\cal C}_\mathrm{Teixeira} \stackrel{\mathrm{def}}{=} \frac{Dq_c}{2q_s(1-\widehat{\mathrm{RH}})K} \left(-1 + \sqrt{ 1+ \frac{4q_s(1-\widehat{\mathrm{RH}}) K}{Dq_c} } \right),
\end{equation}
where $\widehat{\mathrm{RH}} \stackrel{\mathrm{def}}{=} \min\{\mathrm{RH}, 1-10^{-9}\}$ bounds relative humidity to $1-10^{-9}$ to ensure reasonable asymptotics, $q_s = q_s(T, p)$ is the saturation specific humidity \cite{lohmann2016introduction}, and $\{D, K\}$ are the detrainment rate and the erosion coefficient, which are the two tuning parameters of the Teixeira scheme. 

Besides those three traditional schemes, we additionally train the three neural networks (cell-, neighborhood-, and column-based NNs) from \citeA{grundner2022} on the DYAMOND data. These three NNs receive their inputs either from the same grid cell, the vertical neighborhood of the grid cell, or the entire grid column. Thus, they differ in the amount of vertical locality that is assumed for cloud cover parameterization. As the `undersampling step' has to be done at a cell-based level, we omit it when pre-processing the training data for the column-based NN. Nevertheless, the column-based NN is evaluated on the same validation set as all other models.

Now that we have introduced three semi-empirical cloud cover schemes, which can be used as baselines, we are ready to derive a hierarchy of data-driven cloud cover schemes. 

\subsection{Developing Parsimonious Models via Sequential Feature Selection}
Our goal is to develop parameterizations for cloud cover that are not only performant, but also simple and interpretable. Providing many, possibly correlated features to a model may needlessly increase its complexity and allow the model to learn spurious links between its inputs and outputs \cite{nowack2020causal}, impeding both interpretability \cite{molnar2020} and generalizability \cite{brunton2016discovering}. Therefore, we instead seek parsimonious models. As our feature selection algorithm we use (forward) sequential feature selection (SFS). 

\subsubsection{Sequential Feature Selection}
SFS starts without any features and carefully selects and adds features to a given type of model (e.g., a second-order polynomial) in a sequential manner. At each iteration, SFS selects the feature that optimizes the model's performance on a computationally feasible subset of the training set, which is sufficiently large to ensure robustness (see also Sec~\ref{sec:data_dyamond}). More specifically; let $\mathcal{F}$ contain all potential features of a model (type) $M$. Let us further assume that the SFS approach has already chosen $n$ features $P_n \subseteq \mathcal{F}$ at a given iteration (note that $P_0 := \emptyset$). In the next iteration, the SFS method adds another feature $P_{n+1} = P_n \cup \{\hat{f}\}$, such that $\hat{f} \in \mathcal{F}\setminus P_n$ maximizes the model's performance as measured by the $R^2$-value. Thus, the SFS method tests whether
\begin{equation*}
R^2(M_{P_n \cup \{\hat{f}\}}) \geq R^2(M_{P_n \cup \{\hat{g}\}})
\end{equation*}
indeed holds on the training subset for all features $\hat{g} \in \mathcal{F}\setminus P_n$. With the SFS approach, we discourage the choice of correlated features and enforce sparsity by selecting a controlled number of features that already lead to the desired performance. However, if two highly correlated features are both valuable predictors (as will be the case with $\mathrm{RH}$ and $\partial_z\mathrm{RH}$), the SFS NN would pick them nonetheless. Another benefit is that by studying the order of selected variables, optionally with the corresponding performance gains, we can gather intuition and physical knowledge about the task at hand. On the way, we will obtain an approximation of the best-performing set of features for a given number of features. There is however no guarantee of it truly being the best-performing feature set due to the greedy nature of the feature selection algorithm, which decreases its computational cost. Due to the high cost, we could only verify that the models would pick the same first two features (or four features in the case of the linear model) using a non-greedy selector. However, we found that for some random data subsets the second-order polynomial temporarily outperforms the third-order polynomial due to the earlier pick of a third-order feature that decreased the score later on.

\subsubsection{Linear Models and Polynomials}
\label{sec:pol}
% ~/workspace_icon-ml/symbolic_regression/baselines/multivariate_polynomial_fit_v2.ipynb
% ~/workspace_icon-ml/symbolic_regression/baselines/linear_v2.ipynb
We allow first-order (i.e., linear models), second-order, and third-order polynomials. For each of these model types, we run SFS using the \textit{SequentialFeatureSelector} of scikit-learn \cite{pedregosa2011scikit}. In the case of linear models, the pool of features $\mathcal{F}_1$ to choose from is precisely $\mathcal{F}$ (see Sec~\ref{sec:data_dyamond}). For second-order polynomials, $\mathcal{F}_2$ also includes second-degree monomials of the features in $\mathcal{F}$, i.e.,
\begin{equation*}
    \mathcal{F}_2 = \{xy \, \vert \, x,y \in \mathcal{F}\} \cup \mathcal{F}.
\end{equation*}
Analogously we also consider third-degree monomials 
\begin{equation*}
    \mathcal{F}_3 = \{xyz  \, \vert \, x,y,z \in \mathcal{F}\} \cup \mathcal{F}_2
\end{equation*}
in the case of third-order polynomials. Thus, the set of possible terms grows from $25$ to $325$ for the second-order and would grow to $2925$ for the third-order polynomials. However, to circumvent memory issues for the third-order polynomials, we restrict the pool of possible features to combinations of the ten most important features. The choice of these ten features is informed by the SFS NNs (Sec~\ref{sec:sfsnns}), which are able to select informative features for nonlinear models. In addition to these ten features, we also incorporate air pressure to later classify samples into physically interpretable cloud regimes. To be specific, this implies that
\begin{equation*}
    \mathcal{F}_3 = \{xyz \, \vert \, x,y,z \in \{1, \mathrm{RH}, q_i, q_c, T, \partial_z\mathrm{RH}, \partial_{zz}p, \partial_zp, \partial_{zz}\mathrm{RH}, \partial_zT, p_s, p\}\}.
\end{equation*}
By considering combinations of only eleven features, we reduce the total amount of possible terms from $2925$ to $364$. After obtaining sequences of selected features for each of the three model types, we fit sequences of models with up to ten features each using ordinary least squares linear regression.

\subsubsection{Neural Networks} 
\label{sec:sfsnns}
We train a sequence of SFS NNs with up to ten features using the ``mlxtend'' Python package \cite{raschkas_2018_mlxtend}. As in the case of the linear models, the pool of possible features is $\mathcal{F}$. We additionally train an NN with all $24$ features in $\mathcal{F}$ for comparison purposes. As our regression task is similar in nature (including the vertical locality assumptions it makes for the features), we use the ``Q3 NN'' model architecture from \citeA{grundner2022} for all SFS NNs. ``Q3 NN'''s architecture has three hidden layers with $64$ units each; it uses batch normalization and its loss function includes $L^1$ and $L^2$-regularization terms following hyperparameter optimization. After deriving the sequence of ten features on small training data subsets (see Sec~\ref{sec:feat_ranking}) we train the final SFS NNs on the entire training data set, always limiting the number of training epochs to $25$ and making use of early stopping. Without the greedy assumption of the SFS approach we would already need to test more than $2000$ NNs for three features. 

Due to the flexibility of NNs, when combining SFS with NNs, we obtain a sequence of features that is not bound to a particular model structure. In Sec~\ref{sec:pol} and \ref{sec:symb_fit}, we therefore reuse the SFS NN feature rankings for other nonlinear models to restrict their set of possible features. The combination of SFS with NNs also yields a tentative upper bound on the accuracy one can achieve with $N$ features: If we assume that i) SFS provides the best set of features for a given number of features $N$; and ii) the NNs are able to outperform all other models given their features, one would not be able to outperform the SFS NNs with the same number of features. Even though the assumptions are only met approximately, we still receive helpful upper bounds on the performance of any model with $N$ features.

\subsection{Symbolic Regression Fits}
\label{sec:symb_fit}
% Why do I use symbolic regression models?
To improve upon the analytical models of Sec~\ref{sec:ex_schemes} and \ref{sec:pol} without compromising interpretability, we use recently-developed symbolic regression packages.
% Which ones do I use? Advantages.
We choose the PySR \cite{pysr} and the default GP-GOMEA \cite{virgolin2021improving} libraries, which are both based on genetic programming. GP-GOMEA is one of the best symbolic regression libraries according to SRBench, a symbolic regression benchmarking project that compared 14 contemporary symbolic regression methods \cite{cava2021}. PySR is a very flexible, efficient, well-documented, and well-maintained library. In PySR, we choose a large number of potential operators to enable a wide range of functions (see \ref{app:pysr} for details).
% Which ones did I try, why did they not work?
We also tried AIFeynman and found that its underlying assumption that one could learn from the NN gradient was problematic for less idealized data. Other promising packages from the SRBench competition, such as DSR/DSO and (Py)Operon, are left for future work.
% Number of features - problematic in symbolic regression tools
PySR and GP-GOMEA can only utilize a very limited number of features. Regardless of the number of features we provide, GP-GOMEA only uses $3-4$, while PySR uses $5-6$ features. For this reason, PySR also has a built-in tree-based feature selection method to reduce the number of potential features. Since the SFS NNs from Sec~\ref{sec:sfsnns} already provide a sequence of features that can be used in general, nonlinear cases, we instead select the first five of these features to maximize comparability between models. The decision to run PySR with five features is also motivated by the good performance ($R^2 > 0.95$) of the corresponding SFS NN (see Sec~\ref{sec:perfcompl}).
% Criteria for selecting which PySR- and GP-GOMEA fits to show.
Each run of the PySR or GP-GOMEA algorithms adds new candidates to the list of final equations. From $\approx\,600$ of resulting equations, we select those that have a good skill ($R^2 > 0.9$), are interpretable, and satisfy most of the physical constraints that we define in the following section. The search itself is performed on the normalized training data (see also Sec~\ref{sec:data_dyamond}). As a final step, we refine the free parameters in the equation using the Nelder-Mead and BFGS optimizers (as in Sec~\ref{sec:ex_schemes}).

\section{Model Evaluation \label{sec:model_eval}}
\subsection{Physical Constraints}
\label{sec:phys_cons}
To facilitate their use, we postulate that simple equations for cloud cover ${\cal C}(X)$ ought to satisfy certain physical constraints \cite{gentine2021deep, kashinath2021physics}: 1) The cloud cover output should be between $0$ and $100\%$; 2) an absence of cloud condensates should imply an absence of clouds; 3-5) when relative humidity or the cloud water/ice mixing ratios increase (keeping all other features fixed), then cloud cover should not decrease; 6) cloud cover should not increase when temperature increases; 7) the function should be smooth on the entire domain. We can mathematically formalize these physical constraints (PC):
\begin{enumerate}[1)]
    \item PC$_1$: ${\cal C}(X) \in [0, 100]\%$
    \item PC$_2$: $(q_c, q_i) = 0 \Rightarrow {\cal C}(X) = 0$   
    \item PC$_3$: $\partial {\cal C}(X)/\partial \mathrm{RH} \geq 0$
    \item PC$_4$: $\partial {\cal C}(X)/\partial q_c \geq 0$
    \item PC$_5$: $\partial {\cal C}(X)/\partial q_i \geq 0$
    \item PC$_6$: $\partial {\cal C}(X)/\partial T\leq 0$
    \item PC$_7$: ${\cal C}(X)$ is a smooth function
\end{enumerate}
While these physical constraints are intuitive, they will not be respected by data-driven cloud cover schemes if they are not satisfied in the data. In the DYAMOND data, the first physical constraint is always satisfied, and PC$_2$ is satisfied in $99.7\%$ of all condensate-free samples. The remaining $0.3\%$ are due to noise induced during coarse-graining. In order to check whether PC$_3$\textendash{}PC$_6$ are satisfied in our subset of the coarse-grained DYAMOND data, we extract $\{q_c, q_i, \mathrm{RH}, T\}$. We then separate the variable whose partial derivative we are interested in. Bounded by the min/max-values of the remaining three variables, we define a cube in this three-dimensional space, which we divide into $N^3$ equally-sized cubes. In this way, the three variables of the samples within the cubes become more similar with increasing $N$. If we now fit a linear function in a given cube with the separated variable as the inputs and cloud cover as the output, then we can use the sign of the function's slope to know whether the physical constraint is satisfied.

On one hand, the test is more expressive the smaller the cubes are, as the samples have more similar values for three of the four chosen variables and we can better approximate the partial derivative with respect to the separated variable. However, we only guarantee similarity in three variables (omitting, e.g., pressure). On the other hand, as the size of the cubes decreases, so does the number of samples contained in a cube, and noisy samples may skew the results. We therefore only consider the cubes that contain a sufficiently large number of samples (at least $10^4$ out of the $2.9\cdot 10^8$). 

\begin{table}[!htb]
\centering
\caption{The percentage of data cubes that fulfill a given physical constraint. Only the cubes with a sufficiently large amount of samples are taken into account. The last column shows the proportion of cubes (across all sizes we consider) in which the constraint is satisfied on average.}
\rowcolors{2}{}{blue!5}
\label{tab:phys_cons}
\begin{tabular}{l c c c c c c c c}
& \multicolumn{7}{c}{\textbf{(Maximum) Number of data cubes}} & \\
\cmidrule{2-8}
 & 1 & $2^3$ & $3^3$ & $4^3$ & $5^3$ & $6^3$ & $7^3$ & Average (\%) \\
 \hline
 \textbf{PC$_3$} & $100$ & $100$ & $100$ & $100$ & $100$ & $100$ & $100$ & $100$ \\
 \textbf{PC$_4$} & $100$ & $100$ & $83$ & $90$ & $73$ & $78$ & $71$ & $77.5$ \\
 \textbf{PC$_5$} & $100$ & $100$ & $85$ & $50$ & $81$ & $83$ & $68$ & $73.8$ \\
 \textbf{PC$_6$} & $100$ & $50$ & $100$ & $67$ & $72$ & $89$ & $75$ & $77.7$ 
\end{tabular}
\end{table}

We collect the results in Table \ref{tab:phys_cons}, and find that the physical constraint PC$_3$ (with respect to RH) is always satisfied. The other constraints are satisfied in most (on average $76 \%$) of the cubes. Thus, from the data we can deduce that the final cloud cover scheme should satisfy PC$_1$\textendash{}PC$_3$ in all and PC$_4$\textendash{}PC$_6$ in most of the cases.

To enforce PC$_1$, we always constrain the output to $[0,100]$ before computing the MSE. With the exception of the linear and polynomial SFS models, we already ensure PC$_1$ during training. For PC$_2$, we can define cloud cover to be $0$ if the grid cell is condensate-free. We can combine PC$_1$ and PC$_2$ to define cloud fraction $\cal{C} $ (in \%) as
\begin{equation}
\label{eq:pc12}
{\cal C}(X)=\begin{cases}
			0, & \text{if $q_i + q_c = 0$}\\
            100 \cdot \max\{\min\{f(X), 1\}, 0\}, & \text{otherwise},
		 \end{cases}
\end{equation}
and our goal is to learn the best fit for $f(X)$. In the case of the Xu-Randall and Teixeira schemes, ensuring PC$_2$ is not necessary since they satisfy the constraint by design. \\

\subsection{Performance Metrics}
We use different metrics to train and validate the cloud cover schemes. We always train to minimize the mean squared error (MSE), which directly measures the average squared mismatch of the predictions $f(x_i)$ (usually set to be in $[0,100]\%$) and the corresponding true (cloud cover) values $y_i$:
\begin{equation}
\text{MSE} \stackrel{\mathrm{def}}{=} \frac1N \sum_{i=1}^N ({\cal C}(x_i) - y_i)^2.
\end{equation}
The coefficient of determination $R^2$-value takes the variance of the output $Y = \{y_i\}_{i=1}^N$ into account:
\begin{equation}
R^2 \stackrel{\mathrm{def}}{=} 1 - \frac{\text{MSE}}{\text{Var}(Y)}.
\end{equation}
To compare discrete univariate probability distributions $P$ and $Q$, we use the Hellinger distance
\begin{equation}
H(P, Q) \stackrel{\mathrm{def}}{=} \frac1{\sqrt{2}} \lVert \sqrt{P} - \sqrt{Q} \rVert_2.
\end{equation}
As opposed to the Kullback-Leibler divergence, the Hellinger distance between two distributions is always symmetric and finite (in $[0, 1]$).

As our measure of complexity we use the number of (free/tunable/trainable) parameters of a model. A clear limitation of this complexity measure is that, e.g., the expression $f(x) = ax$ is considered as complex as $g(x) = \sin(\exp(ax))$. However, in this study, most of our models (i.e., the linear models, polynomials, and NNs) do not contain these types of nested operators. Instead, each additional parameter usually corresponds to an additional term in the equation. In the case of symbolic regression tools, operators are already taken into account (see \ref{app:pysr}) during the selection process, and we find that the number of trainable parameters suffices to compare the complexity of our symbolic equations in their simplified forms. Finally, this complexity measure is one of the few that can be used for both analytical equations and NNs.

\subsection{Cloud Regime-Based Evaluation} 
\label{sec:cl_reg}
We define four cloud regimes based on air pressure $p$ and the total cloud condensate $q_t$ (cloud water plus cloud ice) mixing ratio:
\begin{enumerate}
    \item Low air pressure, little condensate (cirrus-type cloud regime)
    \item High air pressure, little condensate (cumulus-type cloud regime)
    \item Low air pressure, substantial condensate (deep convective-type cloud regime)
    \item High air pressure, substantial condensate (stratus-type cloud regime)
\end{enumerate}
Pressure or condensate values that are above their medians ($78\,787$ Pa and $1.62\cdot 10^{-5}$ kg/kg) are considered to be large, while values below the median are considered small. Each regime has a similar amount of samples (between $35$ and $60$ million samples per regime). In this simplified data split, based on \citeA{rossow1991}, air pressure and total cloud condensate mixing ratio serve as proxies for cloud top pressure and cloud optical thickness. These regimes will help decompose model error to better understand the strengths and weaknesses of each model, discussed in the following section.

\section{Results}
\subsection{Performance on the Storm-Resolving (DYAMOND) Training Set}
\label{sec:results_dyamond}
In this section, we train the models we introduced in Sec~\ref{sec:model_families} on the (coarse-grained) DYAMOND training data and compare their performance and complexity on the DYAMOND validation data. We start with the sequential feature selection's results.

\subsubsection{Feature Ranking}
\label{sec:feat_ranking}
% SFS
% workspace_icon-ml/symbolic_regression/baselines/sfs_order_of_vars.ipynb
We perform 10 SFS runs for each linear model, polynomial, and NN from Sec~\ref{sec:symb_fit}. Each run varies the random training subset, which consists of $\mathcal{O}(10^5)$ samples in the case of NNs and $\mathcal{O}(10^6)$ samples in the case of polynomials (as polynomials are faster to train). We then average the rank of a selected feature and note it down in brackets. We omit the average rank if it is the same for each random subset. By $\mathcal{P}_d, d \in \{1,2,3\}$ we denote polynomials of degree $d$ (e.g., $\mathcal{P}_1$ groups linear models). The sequences in which the features are selected are:
% To test how a removal of condensate-free cells would affect the order in which features are selected by NNs, we introduce NNs$^{\ast}$, which learn the SFS features from the entire dataset.
\begin{align*}
\text{$\mathcal{P}_1$: }& \mathrm{RH} \rightarrow T \rightarrow \partial_z\mathrm{RH} \rightarrow q_i [4.3] \rightarrow \partial_{zz}p [4.7] \rightarrow q_c \rightarrow U \rightarrow \partial_{zz}q_c \rightarrow \partial_{z} q_v \rightarrow z_g  \\
\text{$\mathcal{P}_2$: } & \mathrm{RH} \rightarrow T \rightarrow q_c q_i \rightarrow \mathrm{RH} \partial_z\mathrm{RH} \rightarrow T \partial_z\mathrm{RH} [5.6] \rightarrow q_v \mathrm{RH} [6.4] \rightarrow T \mathrm{RH} [7.4] \rightarrow \\ 
& \mathrm{RH}^2 [7.9] \rightarrow \partial_zq_v [9.2] \rightarrow U [10.1] \\
\text{$\mathcal{P}_3$: } & \mathrm{RH} \rightarrow T \rightarrow q_c q_i \rightarrow T^2 \mathrm{RH} [4.4] \rightarrow \mathrm{RH}^2 [5.4] \rightarrow T^2 [6.7] \rightarrow \mathrm{RH} \partial_z\mathrm{RH} [7.4] \rightarrow \\ 
& \partial_z\mathrm{RH} [8.3] \rightarrow p^2\partial_{zz}p  [8.8] \rightarrow T \partial_z \mathrm{RH} [9.4] \\
\text{NNs: } & \mathrm{RH} \rightarrow q_i \rightarrow q_c \rightarrow T [4.1] \rightarrow \partial_z\mathrm{RH} [4.9] \rightarrow \partial_{zz}p [6.7] \rightarrow \partial_{z}p [8.1] \rightarrow \\
& \partial_{zz} \mathrm{RH} [8.3] \rightarrow \partial_zT [10.0] \rightarrow p_s [10.1] \\
% \text{NNs$^{\ast}$: } & \mathrm{RH} \rightarrow q_i \rightarrow q_c \rightarrow T [4.1] \rightarrow \partial_z\mathrm{RH} [4.9] \rightarrow \partial_{zz}p [7.1] \rightarrow \partial_{zz}\mathrm{RH} [8.7] \rightarrow \\ 
% & \partial_z p [9.0] \rightarrow p_s [9.3] \rightarrow \partial_zT [9.5] \\
\end{align*}

% Analysis of the SFS results
Regardless of the model, the selection algorithm chooses RH as the most informative feature for predicting cloud cover. This is consistent with, e.g., \citeA{walcek1994cloud}, who considers RH to be the best single indicator of cloud cover in most of the troposphere. Considering that the cloud cover in the high-resolution data was only derived from the cloud condensate mixing ratio, the models' prioritization of RH is quite remarkable. 
From the feature sequences, we can also deduce that cloud cover depends on the mixing ratios of cloud condensates in a very nonlinear way: The polynomials choose $q_i q_c$ as their third feature and do not use any other terms containing $q_i$ or $q_c$. The NNs choose $q_i$ and $q_c$ as their second and third features, and are able to express a nonlinear function of these two features. The linear model cannot fully exploit $q_i$ and $q_c$ and hence attaches less importance to them. 

% As we can see, cloud water and ice remain important predictors even without considering condensate-free, and thus cloud-free cells. Therefore, there is little difference between the sequences for NNs and NNs$^{\ast}$ (the first six features are the same).

Since $\mathrm{RH}$ and $T$ are chosen as the most informative features for the linear model, we can derive a notable linear dependence of cloud cover on these two features (the corresponding model being $f(\mathrm{RH}, T) = 41.31\mathrm{RH}-15.54T+44.63$). However, given the possibility, higher order terms of $T$ and $\mathrm{RH}$ are chosen as additional predictors over, for instance, $p$ or $q_v$. Finally, $\partial_z \mathrm{RH}$ is an important recurrent feature for all models. Depending on the model, the coefficient associated with $\partial_z \mathrm{RH}$ can be either negative or positive. If $\partial_z \mathrm{RH} \neq 0$, one can assume some variation of cloud cover (i.e., cloud area fraction) vertically within the grid cell. Thus, $\partial_z\mathrm{RH}$ is a meaningful proxy for the subgrid vertical variability of cloud area fraction. Since the effective cloud area fraction of the entire grid cell is related to the maximum cloud area fraction at a given height within the grid cell, this could explain the significance of $\partial_z\mathrm{RH}$. 

\subsubsection{Balancing Performance and Complexity}
\label{sec:perfcompl}
\begin{figure}
\centering
\hspace*{-4cm}\includegraphics[width=1.5\textwidth]{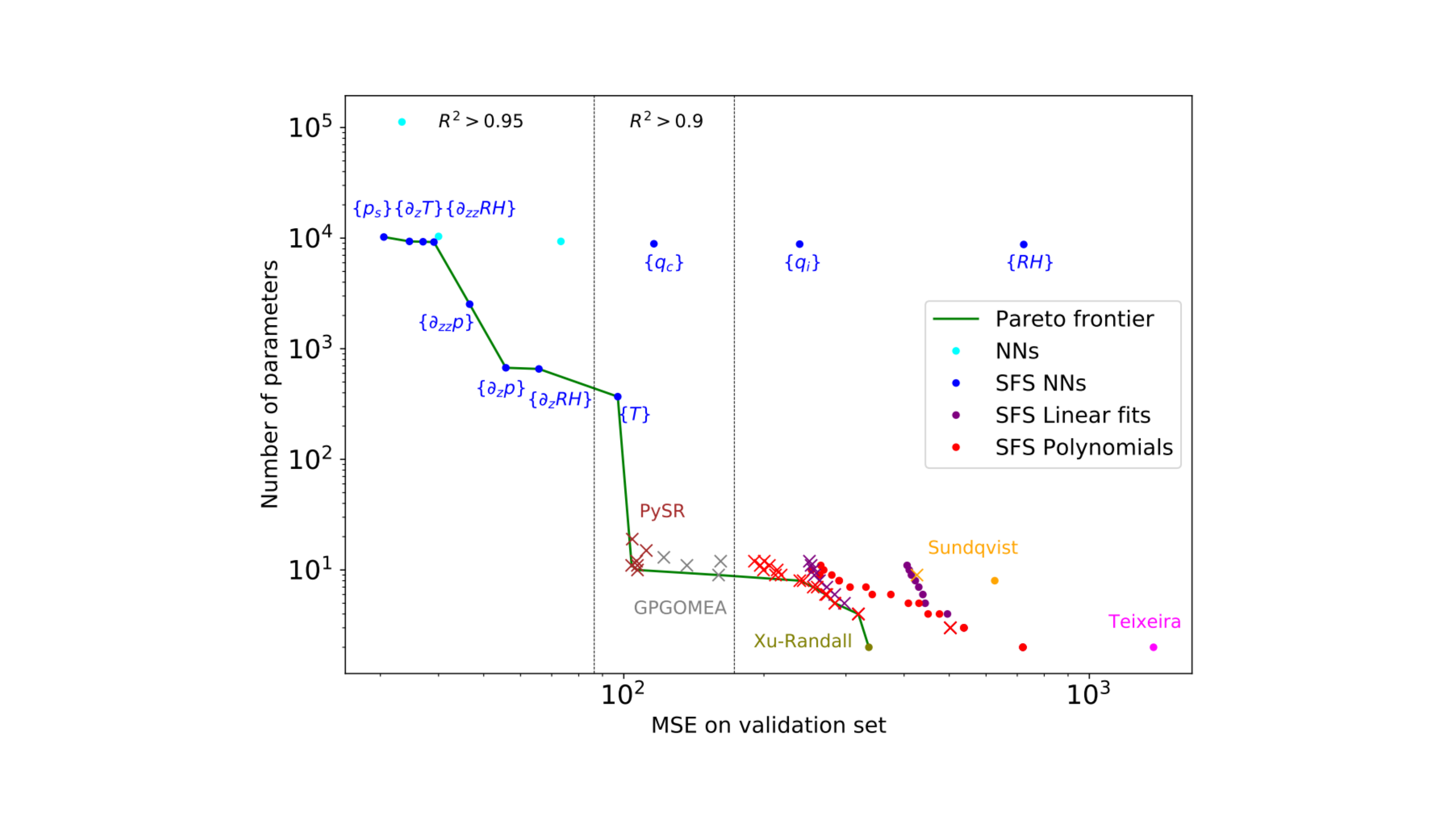}
\caption{All models described in Sec~\ref{sec:model_families} in a performance $\times$ complexity plot. The dashed vertical lines mark the $R^2 = 0.95$- and $R^2 = 0.9$-boundaries. Models marked with a cross satisfy the second physical constraint PC$_2$ (using equation (\ref{eq:pc12})). Only the best PySR and GP-GOMEA symbolic regression fits are shown. The NNs in cyan are the column-, neighborhood- and cell-based NNs when read from left to right. The SFS NN with the lowest MSE contains all $24$ features described in Sec~\ref{sec:data_dyamond}. For the SFS NNs, the last added feature is specified in curly brackets. Since the validation MSE of the SFS NNs decreases with additional features, we can extract the features for a given SFS NN by reading from right to left (e.g., the features of the SFS NN marked with $\left\{q_c\right\}$ are $\{q_i, q_c, \mathrm{RH}\}$).}
\label{fig:pareto_plot}
\end{figure}

% Pareto Plot
In Fig~\ref{fig:pareto_plot}, we depict all of our models in a performance $\times$ complexity plane. We measure performance as the MSE on the validation (sub)set of the DYAMOND data and use the number of free parameters in the model as our complexity metric. We add the Pareto frontier, defined to pass through the best-performing models of a given complexity. The SFS sequences described above are used to train the SFS models of the corresponding type. The only exception is the swapped order of $\partial_zp$ and $\partial_{zz}p$ for the NNs, as we base the sequence shown in Fig~\ref{fig:pareto_plot} on a single SFS run.
% SFS NNs
For the SFS NNs with 4\textendash{}7 features, it was possible to reduce the number of layers and hidden units without significant performance degradation, which reduced the number of free parameters by about an order of magnitude and put them on the Pareto frontier.

% PC_1 and PC_2
For most models, we train a second version that does not need to learn that condensate-free cells are always cloud-free, but for which the constraint is embedded by equation (\ref{eq:pc12}). For such models, condensate-free cells are removed from the training set. In addition to the schemes of Xu-Randall and Teixeira (see Sec~\ref{sec:ex_schemes}), we find that it is also not necessary to enforce PC$_2$ in the case of NNs, since they are able to learn PC$_2$ without degrading their performance. PC$_1$ is always enforced by default for all models.

% Pareto frontier
We find that, even though the Sundqvist and Teixeira schemes are also tuned to the training set, linear models of the same complexity outperform them. However, these linear models do not lie on the Pareto frontier either. The lower performance of the Teixeira scheme is most likely due to the fact that it was developed for subtropical boundary layer clouds. Its MSE experiences a reduction (to $290 \,(\%)^2$) when evaluated exclusively within the subtropics (from 23.4 to 35 degrees north and south). Among the existing schemes, only the Xu-Randall scheme with its two tuning parameters set to $\{\alpha, \beta\} = \{0.9, 9 \cdot 10^5\}$ is on the Pareto frontier as the simplest model. With relatively large values for $\alpha$ and $\beta$, cloud cover is always approximately equal to relative humidity (i.e., ${\cal C} \approx \, \mathrm{RH}^{0.9}$) when cloud condensates are present. The next models on the Pareto frontier are third-order SFS polynomials $\mathcal{P}_3$ with 2\textendash{}6 features with PC$_2$ enforced. To account for the bias term and the output of the polynomial being set to zero in condensate-free cells, the number of their parameters is the number of features plus 2. We then pass the line with $R^2 = 0.9$ and find three symbolic regression fits on the Pareto frontier, each trained on the five most informative features for the SFS NNs. All symbolic regression equations that appear in the plot are listed in \ref{app:eqns}. We will analyze the PySR equation with arguably the best tradeoff between complexity ($11$ free parameters when phrased in terms of normalized variables) and performance ($MSE = 103.95 \,(\%)^2$, improved spatial distribution as illustrated in Fig~S2) in Sec~\ref{sec:pysr_phys_int}. The remaining models on the Pareto frontier are SFS NNs with 4\textendash{}10 features and finally the NN with all 24 features defined in Sec~\ref{sec:data_dyamond} included ($MSE = 30.51 \,(\%)^2$).

% Column-based outperformed by derivatives
Interestingly, the (quasi-local) 24-feature NN is able to achieve a slightly lower MSE ($30.51 \,(\%)^2$) than the (non-local) column-based NN ($33.37 \,(\%)^2$) with its 163 features. The two aspects that benefit the 24-feature NN are the additional information on the horizontal wind speed $U$ and its derivatives, and the smaller number of condensate-free cells in its training set due to undersampling (Sec~\ref{sec:data_dyamond} and \ref{sec:ex_schemes}). The SFS NN with 10 features already shows very similar performance ($MSE =34.64 \,(\%)^2$) to the column-based NN with a (12 times) smaller complexity and fewer, more commonly accessible features.

% Linear vs. Polynomial
Comparing the small improvements of the linear SFS models (up to $MSE = 250.43 \,(\%)^2$) with the larger improvements of SFS polynomials (up to $MSE = 190.78 \,(\%)^2$) with increasing complexity, it can be deduced that it is beneficial to include nonlinear terms instead of additional features in a linear model. For example, NNs require only three features to predict cloud cover reasonably well ($R^2 = 0.933$), and five features are sufficient to produce an excellent model ($R^2 = 0.962$) because they learn to nonlinearly transform these features.

% Other conclusions?
The PySR equations can estimate cloud cover very well ($R^2 \in [0.935, 0.940]$). However, while the PySR equations depend on five features, the NNs are able to outperform them with as few as four features ($R^2 = 0.944$). This suggests that the NNs learn better functional dependencies than PySR, as they do better with less information. However, the improved performance of the NNs comes at the cost of additional complexity and greatly reduced interpretability.

\subsection{Split by Cloud Regimes \label{sec:cloud_regimes}}
\begin{figure}
\centering
\hspace*{-3.5cm}\includegraphics[width=1.5\textwidth]{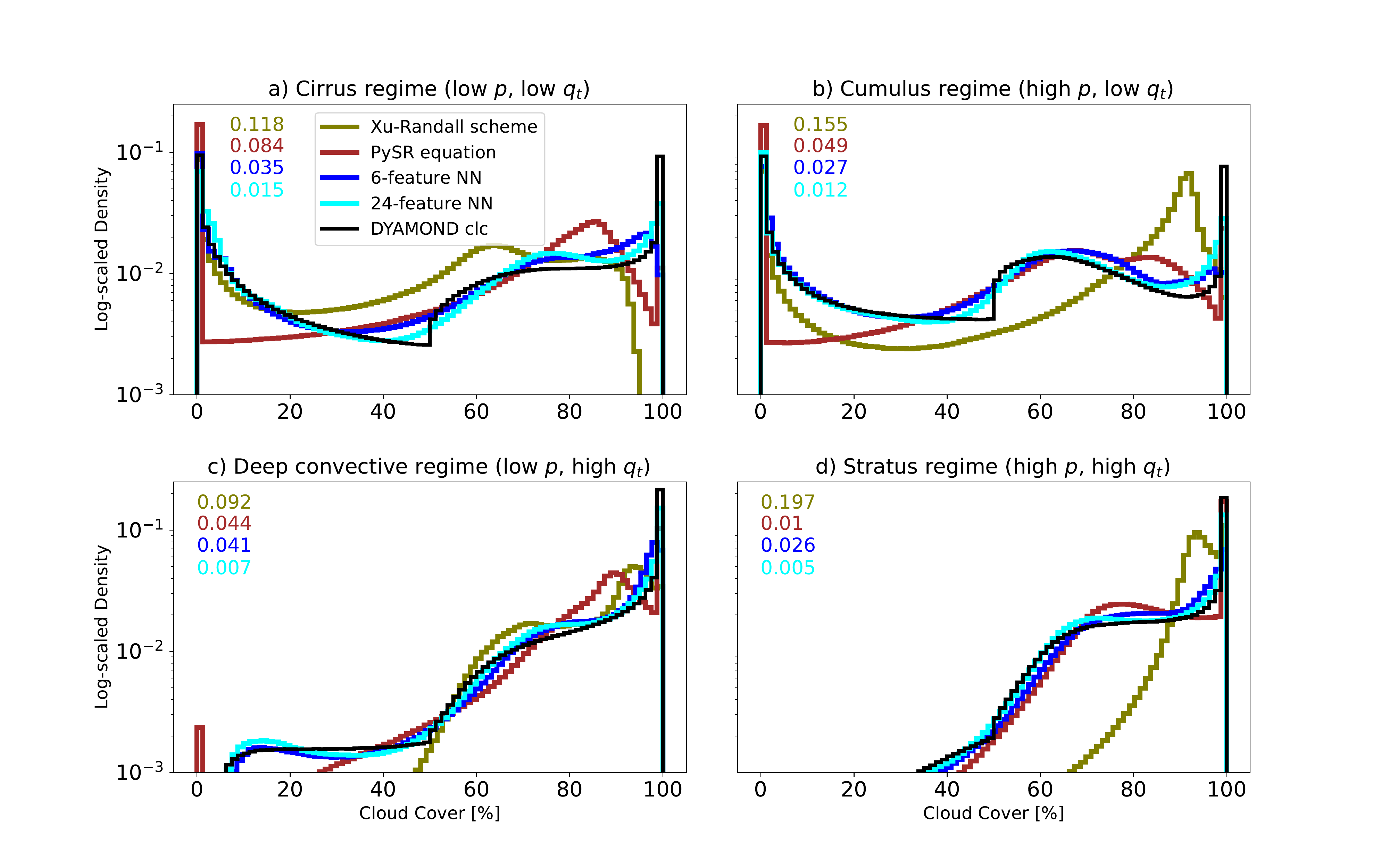}
\caption{Predicted cloud cover distributions of selected Pareto-optimal models evaluated on the DYAMOND data, divided into four different cloud regimes. The numbers in the upper left indicate the Hellinger distance between the predicted and the actual cloud cover distributions for each model and cloud regime.}
\label{fig:hellinger}
\end{figure}

In this section, we divide the DYAMOND data set into the four cloud regimes introduced in Sec~\ref{sec:cl_reg}. In Fig~\ref{fig:hellinger}, we compare the cloud cover predictions of Pareto-optimal models (on Fig~\ref{fig:pareto_plot}'s Pareto frontier) with the actual cloud cover distribution in these regimes. We evaluate the models located at favorable positions on the Pareto frontier (at the beginning to maximize simplicity, at the end to maximize performance, or on some corners to optimally balance both). Of the two PySR equations, we consider the one with the lowest MSE (as in Sec~\ref{sec:pysr_phys_int} later). Furthermore, we explore benefits that arise from training on each cloud regime separately and whether using a different feature set for each regime could ease the transition between regimes.

% General findings 
In general, we find that the PySR equation (except in the cirrus regime) and the 6-feature NN can reproduce the distributions quite well (Hellinger distances $< 0.05$), while the 24-feature NN shows excellent skill (Hellinger distances $\leq 0.015$). However, all models have difficulty predicting the number of fully cloudy cells in all regimes (especially in the regimes with fewer cloud condensates). 

% On the Xu-Randall scheme
Focusing first on the predictions of the Xu-Randall scheme, we find that the distributions exhibit prominent peaks in each cloud regime. By neglecting the cloud condensate term and equating RH with the regime-based median, we can approximately re-derive these modes of the Xu-Randall cloud cover distributions in each regime using the Xu-Randall equation (\ref{eq:xu-randall}). With our choice of $\alpha = 0.9$, this mode is indeed very close (absolute difference at most $8\%$ cloud cover) to the median relative humidity calculated in each regime. By increasing $\alpha$, we should therefore be able to push the mode above $100\%$ cloud cover and thus remove the spurious peak. However, this comes at the cost of increasing the overall MSE of the Xu-Randall scheme.

% Generally the most difficult regime?
For the PySR equation (and also the 24-feature NN), the cirrus regime distribution is the most difficult to replicate. The Hellinger distances suggest that it is the model's functional form, and not its number of features that limits model performance in the cirrus regime. Indeed, the decrease in the Hellinger distance between the PySR equation and the 6-feature NN is larger ($0.049$) than the decrease between the 6- and the 24-feature NN ($0.02$). Technically, the PySR equation has the same features as the 5-feature and not the 6-feature NN, but the Hellinger distances of these two NNs to the actual cloud cover distribution are almost the same (difference of $0.003$ in the cirrus regime). We want to note here that, while the PySR equation features a large Hellinger distance, it actually achieves its best $R^2$ score ($R^2 = 0.84$) in the cirrus regime as the coefficient of determination takes into account the high variance of cloud cover in the cirrus regime.
% PySR in the cirrus regime
In the condensate-rich regimes, the PySR equation is as good as the 6-feature NN and even able to outperform it on the stratus regime. To improve the PySR scheme further in terms of its predicted cloud cover distributions, and combat its underestimation of cloud cover in the cirrus regime, we now explore the effect of focusing on the regimes individually. By training SFS NNs just like in Sec~\ref{sec:feat_ranking} but now on each cloud regime separately, we find new feature rankings:

% Top 5 features per regime across 10 seeds
\begin{align*}
\text{Cirrus regime: }& q_i \rightarrow \mathrm{RH} \rightarrow T [3.4] \rightarrow \partial_z\mathrm{RH}  \rightarrow \partial_{zz}\mathrm{RH} [6.4] \\
\text{Cumulus regime: } & q_i \rightarrow q_c \rightarrow \mathrm{RH} \rightarrow \partial_z\mathrm{RH} [4.5] \rightarrow \partial_{zz}p [5.1] \\
\text{Deep convective regime: } & \mathrm{RH} \rightarrow T \rightarrow \partial_z\mathrm{RH} \rightarrow p_s [5.5] \rightarrow \partial_{zz}\mathrm{RH} [5.6] \\
\text{Stratus regime: } & \mathrm{RH} \rightarrow \partial_z\mathrm{RH} \rightarrow \partial_{zz}p \rightarrow \partial_{zz}\mathrm{RH} [5.9] \rightarrow q_c [6.3] \\
\end{align*}

% PySR and SI
By rerunning PySR within each regime and allowing its discovered equations to depend on the newly found five most important features, we find equations that are better able to predict the distributions of cloud cover. In the supplementary information (SI), we present one of the equations per regime that strikes a good balance between performance and simplicity and show the predicted distributions of cloud cover.
 
% Condensate-poor regimes
As expected, cloud water is not an informative variable in the cirrus regime (with an average rank of $9.5$). Based on $q_i, \mathrm{RH}$ and $T$ alone, we are able to discover equations that reduce the number of cloud-free predictions and improve the distributions for low cloud cover values (Hellinger distances of $\approx 0.05$). We do not attribute these improvements to new input features, but rather to the ability of the equation to adopt a novel structure. Similarly, the features $q_i, q_c$ and $\mathrm{RH}$ are sufficient to decrease the Hellinger distance from $0.049$ to $0.041$ within the cumulus regime. 

% Condensate-rich regimes
In the condensate-rich regimes (deep convective and stratus), cloud water and/or ice are already present, making the exact amount of cloud condensates less pertinent. By focusing on the three most significant features $\mathrm{RH}, T$ and $\partial_z\mathrm{RH}$, we find equations with an enhanced distribution of cloud cover within the deep convective regime (with Hellinger distances of only $0.02$). The equations specific to the deep convective regime display strong nonlinearity, with the equation selected for the SI including a fourth-order polynomial of relative humidity and temperature. While the five most important features of the stratus regime also differ from the SFS NN features of Sec~\ref{sec:feat_ranking}, we were not able to improve upon the Hellinger value of our single PySR equation through exclusive training within the stratus regime. A notable aspect of the stratus regime is the increased significance of $\partial_z \mathrm{RH}$, which is discussed later (see Sec~\ref{sec:rh_grad_stratocumulus}).

% Nevertheless, the equation is not Pareto-optimal
While the approach of deriving distinct equations tailored to each cloud regime, emphasizing regime-specific features, holds potential for improving predicted cloud cover distributions, the resulting MSE across the entire dataset is lower ($\approx 113 \,(\%)^2$) compared to our chosen single PySR equation ($\approx 104 \,(\%)^2$). Moreover, the number of free parameters increases to $33$, which is three times the count of our single PySR equation. Lastly, formulating distinct equations for each cloud regime requires special attention at the regime boundaries to ensure continuity across the entire domain. Therefore, we henceforth focus on equations that generalize across cloud regimes. 

\subsection{Transferability to Different Climate Model Horizontal Resolutions} \label{sec:transfer_res}

Designing data-driven models that are not specific to a given Earth system model and a given grid is challenging. Therefore, in this section we aim to determine which of our selected Pareto-optimal ML models are most general and transferable. We explore the applicability of our schemes at higher resolutions, nowadays also typical for climate model simulations. 

\begin{figure}
\centering
\includegraphics[width=0.6\textwidth]{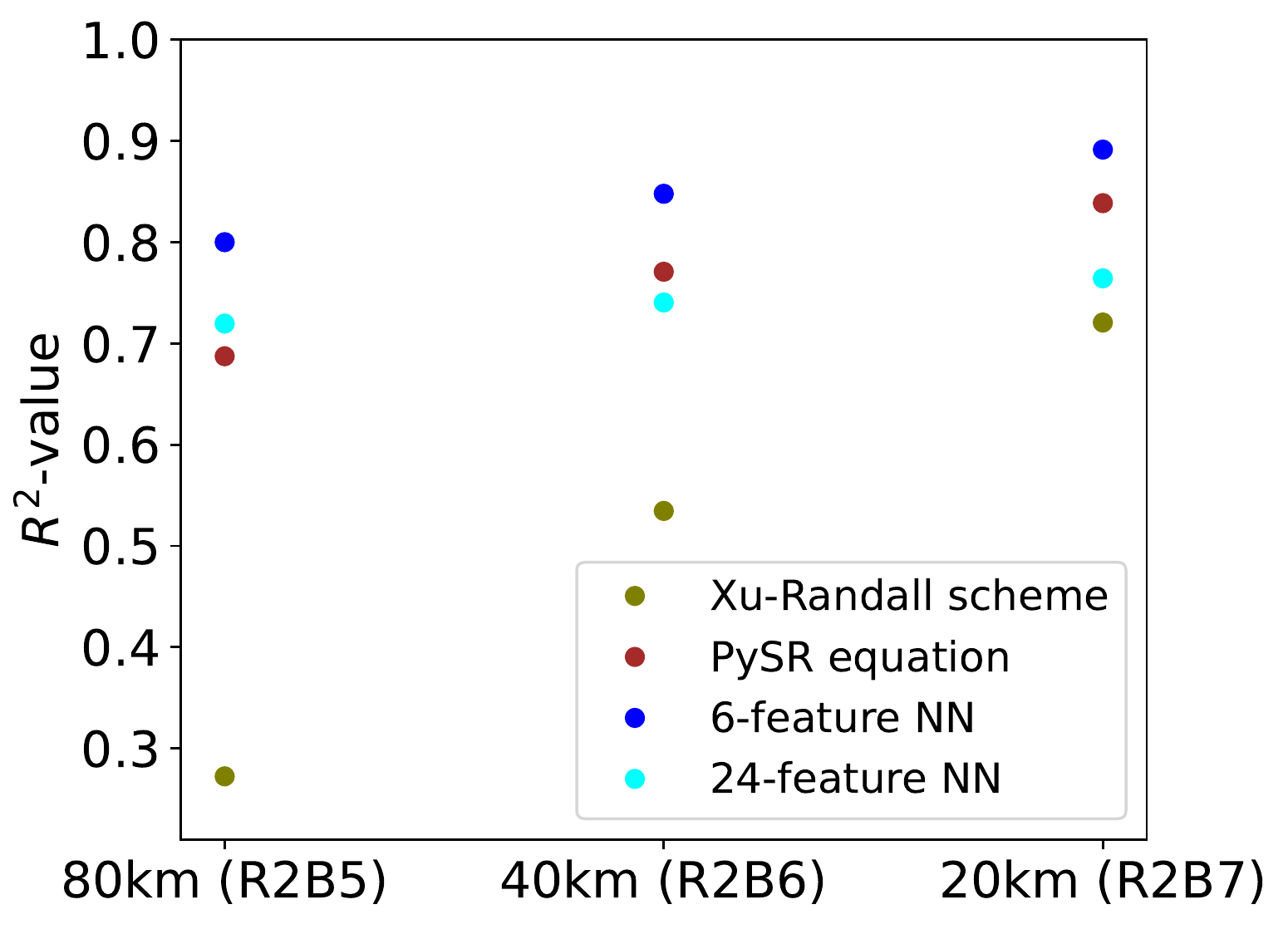}
\caption{Selected Pareto-optimal models evaluated on DYAMOND data (Aug 11\textendash{}20, 2018), coarse-grained horizontally to three different resolutions. Only data below an altitude of $21\,$km is considered.}
\label{fig:transfer_resolutions}
\end{figure}

To evaluate the performance of our models at higher resolutions, we coarse-grain some of the DYAMOND data to horizontal resolutions of $\approx 20\,$km (R2B7) and $\approx 40\,$km (R2B6) to complement our coarse-grained data set at $\approx 80\,$km (R2B5). For simplicity, in this section, we omit any coarse-graining in the vertical and do not retune the schemes for the higher resolutions. In Fig~\ref{fig:transfer_resolutions} we present $R^2$-values for each resolution for the same models as in the previous section. We note that the lack of vertical coarse-graining can explain the slight decrease in performance on $80$\,km when compared to the results depicted in Fig~\ref{fig:pareto_plot}.

We observe a clear, almost linear, tendency of all schemes to improve their $R^2$-score on the coarse-grained data sets as we increase the resolution. The increasing standard deviation $\sigma$ of cloud cover by $\approx 1.6\%$ per doubling of the resolution (with $\sigma \approx 23.8\%$ at $80\,$km) is not sufficient to explain this phenomenon. On the one hand, we find these improvements surprising, considering that the schemes were trained at a resolution of $80\,$km. On the other hand, at the low resolution of $80\,$km, the inputs are averaged over wide horizontal regions and bear very little information about how much cloud cover to expect. At higher resolution, large-scale variables and cloud cover are more closely related. Cloud water and ice reach larger values and become more informative for cloud cover detection. This is evident in the Xu-Randall scheme, which relies heavily on cloud condensates and shows a significant increase in its ability to predict cloud cover at higher resolutions. Our analysis reveals that the most skillful schemes at $20\,$km are the $6$-feature NN and our chosen PySR equation. The 24-feature NN relies on many first- and second-order vertical derivatives in its input, so its deteriorated performance could be an artifact of not vertically coarse-graining the data in this section. 

Overall, the schemes exhibit a noteworthy capacity to be applied at higher resolutions than those used during their training. 

\subsection{Transferability to Meteorological Reanalysis (ERA5) \label{sec:transfer}}

To our knowledge, there is no systematic method to incorporate observations into ML parameterizations for climate modeling. In this section, we take a step towards transferring schemes trained on SRMs to observations by analyzing the ability of the Pareto-optimal schemes to transfer learn the ERA5 meteorological reanalysis from the DYAMOND set. 

\begin{figure}
\centering
\hspace*{-3.4cm}\includegraphics[width=1.5\textwidth]{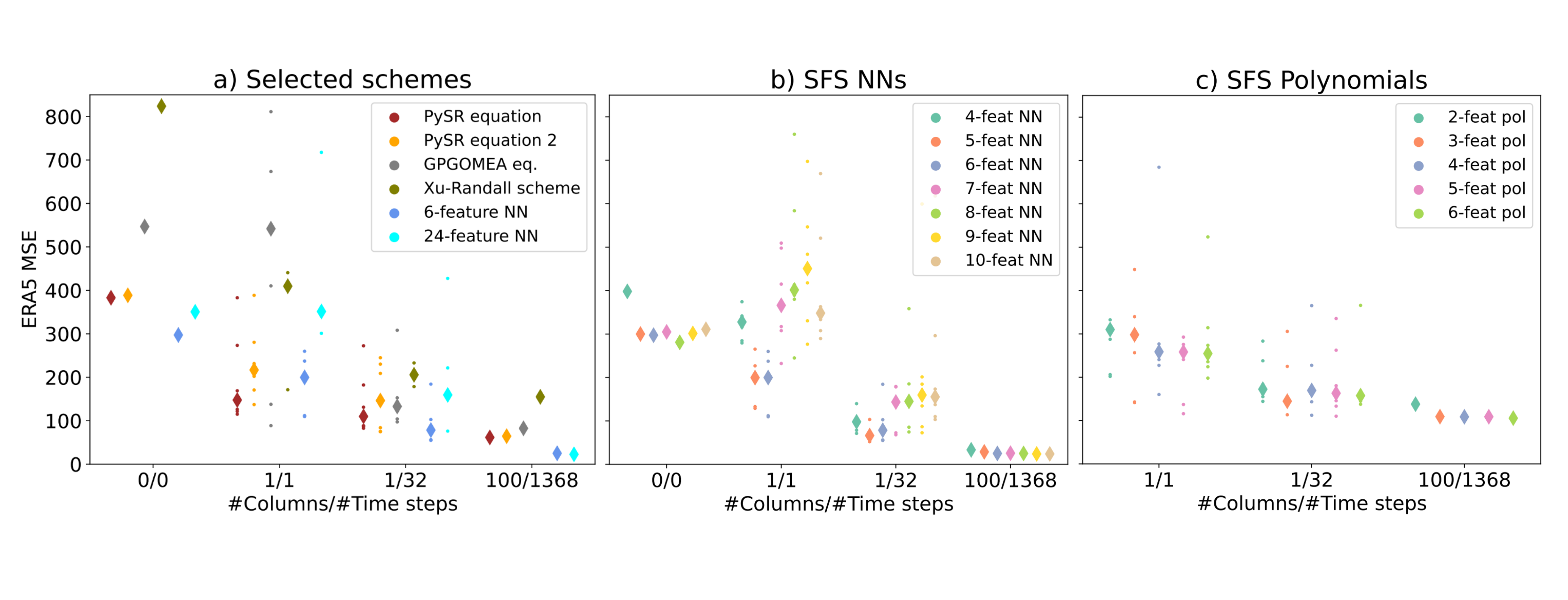}
\caption{Performance of DYAMOND-trained Pareto-optimal cloud cover schemes on the ERA5 data set after transfer learning. The labels on the x-axis denote how many grid columns taken across how many time steps make up the transfer learning training set. Each setting is run with six different random seeds and the diamond-shaped markers indicate the respective medians.}
\label{fig:transfer_learning}
\end{figure}

To do so, we take a certain number (either $1$ or $100$) of random locations, and collect the information from the corresponding grid columns of the ERA5 data over a certain number of time steps in a data set $\mathcal{T}$. Starting from the parameters learned on the DYAMOND data, we retrain the cloud cover schemes on $\mathcal{T}$ and evaluate them on the entire ERA5 data set. In other words, the free parameters of each cloud cover scheme are retuned on $\mathcal{T}$. The retuning method is the same as the original training method, the difference being that the initial model parameters were learned on the DYAMOND data. We can think of $\mathcal{T}$ as mimicking a series of measurements at these random locations, which help the schemes adjust to the unseen data set. Fig~\ref{fig:transfer_learning} shows the MSE of the Pareto-optimal cloud cover schemes on the ERA5 data set after transfer learning on data sets $\mathcal{T}$ of different sizes.

% A bit about the figure
% First column
The first columns of the three panels show no variability because the schemes are applied directly to the ERA5 data without any transfer learning ($\mathcal{T} = \emptyset$). None of the schemes perform well without transfer learning ($R^2 < 0.15$), which is expected given the different distributions of cloud ice and water between the DYAMOND and ERA5 data sets (Fig~\ref{fig:era5}). That being said, the SFS NNs retain their superior performance (MSE $\approx 300 \,(\%)^2$ without retraining), especially compared to the non-retrained SFS polynomials, which exhibit MSEs in the range of $1375 \pm 55 \,(\%)^2$ and are therefore not shown in Panel c. \\
% Second column 
For most schemes, performance increases significantly after seeing one grid column of ERA5 data, with the exception of the SFS NNs with more than $6$ features and the GPGOMEA equation. The performance of the GPGOMEA equation varies greatly between the selected grid columns, and the SFS NNs with many features appear to underfit the small transfer learning training set. The models with the lowest MSEs are (1) the slightly more complex of the two PySR equations (median MSE $= 148 \,(\%)^2$); and (2) the SFS NNs with 5 and 6 features (median MSE $= 200 \,(\%)^2$). While we cannot confirm that fewer features (5-6 features) help with off-the-shelf generalizability of the SFS NNs, they do improve the ability to transfer learn after seeing only a few samples from the ERA5 data.

% Third and fourth column
After increasing the number of time steps to be included in $\mathcal{T}$ to $32$ (corresponding to one year of our preprocessed ERA5 data set), the performances of the models start to converge and the SFS NNs with 5 and 6 features and its large number of trainable parameters outperform the PySR equation (with median $\Delta$MSE $\approx 35 \,(\%)^2$). From the last column we can conclude that a $\mathcal{T}$ consisting of $100$ columns from all available time steps is sufficient for the ERA5 MSE of all schemes to converge. Remarkably, the order from best- to worst-performing model is exactly the same as it was in Fig~\ref{fig:pareto_plot} on the DYAMOND data set (in addition, Fig~S3 visually demonstrates the improved spatial distribution of predicted cloud cover by the fully tuned PySR equation). Thus, we find that the ability to perform well on the DYAMOND data set is directly transferable to the ability to perform well on the ERA5 data set given enough data, despite fundamental differences between the data sets. This suggests a notable degree of structural robustness of the cloud cover models. 

% Ability to transfer learn with few samples
A useful property of a model is that it is able to transfer learn what it learned over an extensive initial dataset after tuning only on a few samples. We can quantify the ability to transfer learn with few samples in two ways: First, we can directly measure the error on the entire data set after the model has seen only a small portion of the data (in our case the ERA5 MSEs of the $1/1$-column). Second, if this error is already close to the minimum possible error of the model, then few samples are really enough for the model to transfer learn to the new data set (in our case, the difference of MSEs in the $1/1$-column and the $100/1368$-column). In terms of the first metric (MSEs in $(\%)^2$), the leading five models are the more complex PySR equation ($147.6$), the 5- and 6-feature NNs ($199.6/199.8$), the simpler PySR equation ($216.8$), and the 6-feature polynomial ($254.6$). In terms of the second metric (difference of MSEs in $(\%)^2$), the top five models are again the more complex PySR equation ($86.0$), the 6-, 5-, and 4-feature polynomials ($149.1/149.4/150.5$), and the simpler PySR equation ($152.3$). If we add both metrics, weighing them equally, then the more complex PySR equation has the lowest inability to transfer learn with few samples ($233.7$), followed by the simpler PySR equation ($369.1$) and the 5- and 6-feature SFS NNs ($370.5/374.5$, where all numbers have units $(\%)^2$). As the more complex PySR equation is leading in both metrics, we can conclude that it is most able to transfer learn after seeing only one column of ERA5 data, and we further investigate its physical behavior in the next section.

\section{Physical Interpretation of the Best Analytical Scheme}
\label{sec:pysr_phys_int}
We find that the two PySR equations on the Pareto frontier (see Fig~ \ref{fig:pareto_plot}) achieve a good compromise between accuracy and simplicity. Both satisfy most of the physical constraints that we defined in Sec~\ref{sec:phys_cons}. In this section, we analyze the (more complex) PySR equation with a lower validation MSE as we showed that it generalized best to ERA5 data (see Fig~\ref{fig:transfer_learning}). We also conclude that the decrease in MSE is substantial enough ($\Delta$MSE $= 3.04 \%^2$) to warrant the analysis of the (one parameter) more complex equation. The equation for the case with condensates can be phrased in terms of physical variables as 
\begin{equation}
\label{best_eq}
f(\mathrm{RH}, T, \partial_z\mathrm{RH}, q_c, q_i) = I_1(\mathrm{RH}, T) + I_2(\partial_z\mathrm{RH}) + I_3(q_c, q_i),
\end{equation}
where
\begin{align*}
I_1(\mathrm{RH}, T) &\stackrel{\mathrm{def}}{=} a_1 + a_2(\mathrm{RH}-\overline{\mathrm{RH}}) + a_3(T-\overline{T}) + \frac{a_4}{2}(\mathrm{RH}-\overline{\mathrm{RH}})^2 + \frac{a_5}{2}(T-\overline{T})^2(\mathrm{RH}-\overline{\mathrm{RH}}) \\
I_2(\partial_z\mathrm{RH}) &\stackrel{\mathrm{def}}{=} a_6^3\left(\partial_z\mathrm{RH} + \frac{3a_7}{2}\right)\left(\partial_z\mathrm{RH}\right)^2 \\
I_3(q_c, q_i) &\stackrel{\mathrm{def}}{=} \frac{-1}{q_c/a_8 + q_i/a_9 + \epsilon}.
\end{align*}
To compute cloud cover in the general case, we plug equation (\ref{best_eq}) into equation (\ref{eq:pc12}), enforcing the first two physical constraints (${\cal C}(X) \in [0, 100]\%$ and in condensate-free cells ${\cal C}(X) = 0$). On the DYAMOND data we find the best values for the coefficients to be
% See symbolic_regression/finding_symmetries/pysr_results_dyamond_on_regimes/optimize_coefs_EQ4_check_physical_eqns.ipynb
\begin{align*}
\{a_1, \dotsc, a_9, \epsilon\} = \{&0.4435, 1.1593, -0.0145\,\mathrm{K}^{-1}, 4.06, 1.3176 \cdot 10^{-3}\,\mathrm{K}^{-2}, \\
&584.8036\,\text{m}, 2\,\text{km}^{-1}, 1.1573\,\text{mg/kg}, 0.3073\,\text{mg/kg}, 1.06\}.
\end{align*}
Additionally, $\overline{\mathrm{RH}} = 0.6025$ and $\overline{T} = 257.06\,\text{K}$ are the average relative humidity and temperature values of our training set. 

In this section, we use our symbolic model to elucidate the fundamental physical components that facilitate the parameterization of cloud cover from storm-resolution data, following the themes outlined in the subsequent subsections.
\begin{figure}
\centering
\hspace*{-3.5cm}\includegraphics[width=1.5\textwidth]{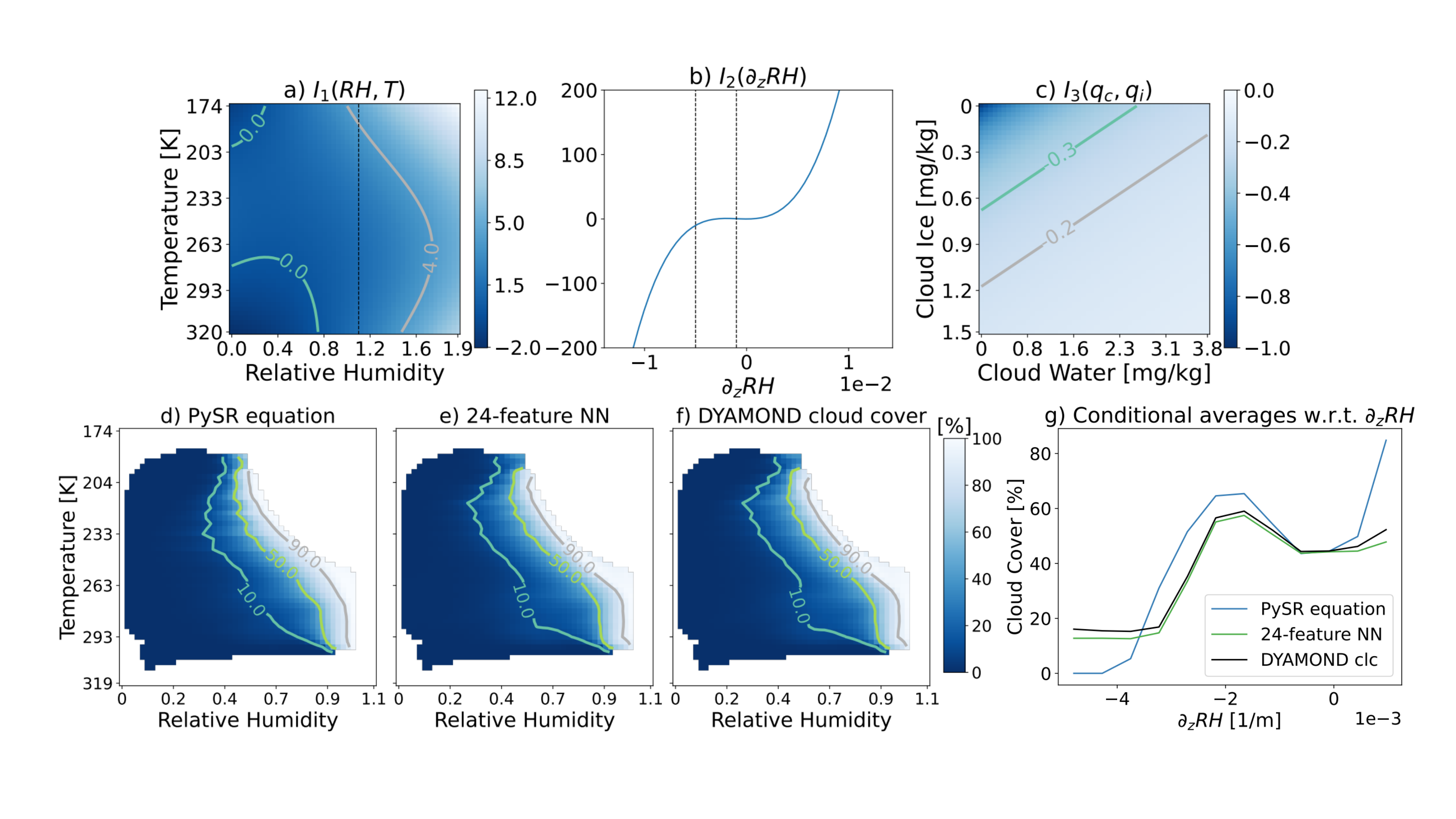}
\caption{Top row: 1D- or 2D-plots of the three terms $I_1, I_2, I_3$ as functions of their inputs. In Panels a and b, the axis-values are bound by the respective minima and maxima in the DYAMOND data set, while those minima/maxima were divided by 5000 in Panel c. The vertical black lines indicate the region of values covered by Panels d\textendash{}g. Bottom row: Conditional average plots of cloud cover with respect to relative humidity and temperature (Panels d\textendash{}f) or $\partial_z\mathrm{RH}$ (Panel g).}
\label{fig:analytic_eq}
\end{figure}

% Discussion of I1, I2, I3 and PCs
\subsection{Relative Humidity and Temperature Drive Cloud Cover, Especially in Condensate-Rich Environments}

\begin{figure}
\centering
\hspace*{-2cm}\includegraphics[width=1.2\textwidth]{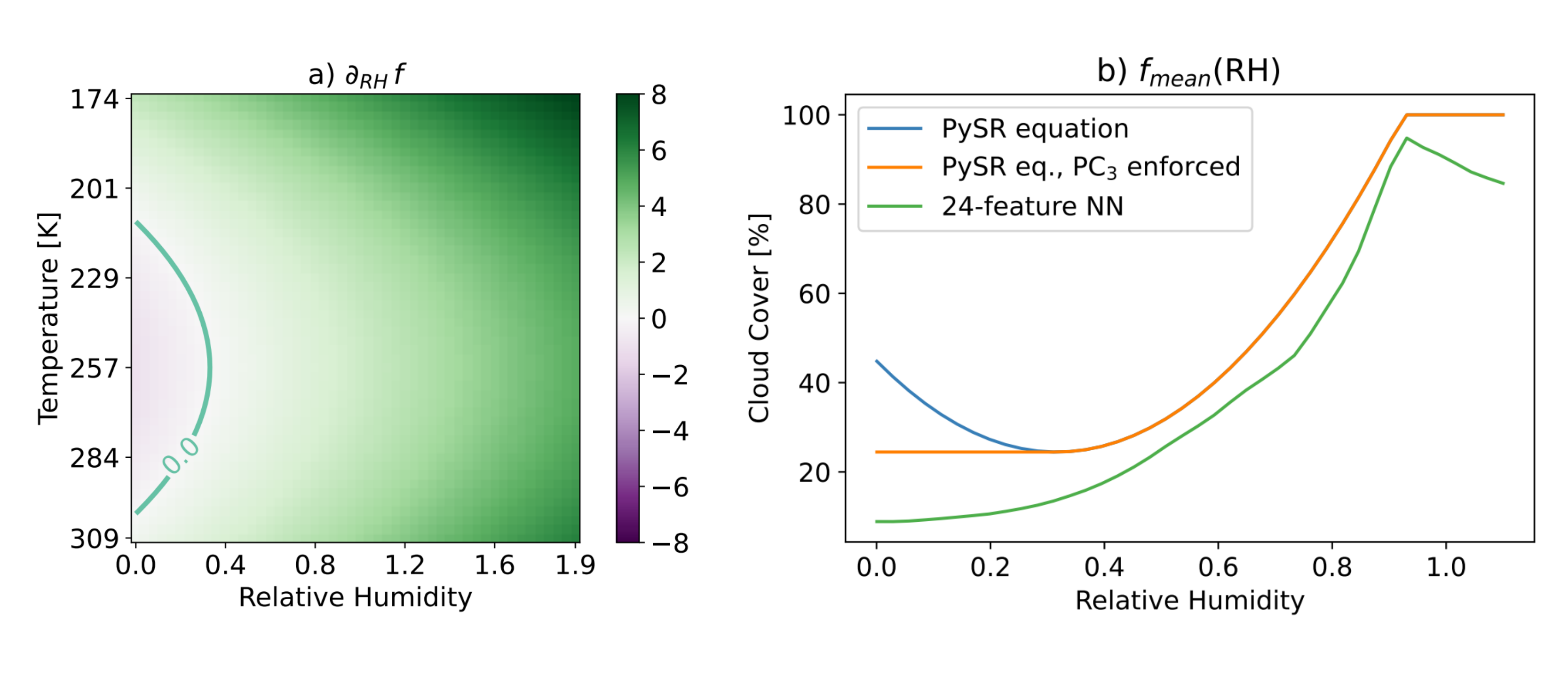}
\caption{Panel a: Contour plot of $\partial_{RH} f$ as a function of relative humidity and temperature. The contour marks the boundary where $\partial_{RH} f = 0$. Panel b: Predictions of the PySR equation (\ref{best_eq}) with and without the modification (\ref{eq_mod}) as a function of relative humidity. For comparison, the predictions of the SFS NN with 24 features are shown. The other features are set to their respective mean values.}
\label{fig:phys_cons_adj}
\end{figure}

The function $I_1(\mathrm{RH}, T)$ can be phrased as a Taylor expansion to third order around the point $(\mathrm{RH}, T) = (\overline{\mathrm{RH}}, \overline{T})$. The first coefficient $a_1$ specifies $I_1$'s contribution to cloud cover for average relative humidity and temperature values, i.e., $a_1 = I_1\left(\overline{\mathrm{RH}}, \overline{T}\right)$. While $\mathcal{C}(X) = a_1$ at $(\overline{\mathrm{RH}}, \overline{T}) $ if $I_2 \approx I_3 \approx 0$, the $I_3$-term dominates when cloud condensates are absent, setting $\mathcal{C}(X)$ to $0$. The following two parameters $a_2$ and $a_3$ are the partial derivatives of equation (\ref{best_eq}) at $(\overline{\mathrm{RH}}, \overline{T})$ w.r.t. relative humidity and temperature, i.e., $a_2 = (\partial I_1/\partial \mathrm{RH})\vert_{(\overline{\mathrm{RH}}, \overline{T})}$ and $a_3 = (\partial I_1/\partial T)\vert_{(\overline{\mathrm{RH}}, \overline{T})}$. As $a_2$ is positive, cloud cover generally increases with relative humidity (see Fig~ \ref{fig:analytic_eq}a and \ref{fig:phys_cons_adj}a). To ensure PC$_3$ ($ \partial {\cal C}/\partial \mathrm{RH}\geq 0$) in all cases, we replace $\mathrm{RH}$ with
\begin{equation}
\label{eq_mod}
\max\{\mathrm{RH}, c_1 - c_2(T-\overline{T})^2\},
\end{equation}
where $c_1 = \overline{\mathrm{RH}} - a_2/a_4 \approx 0.317$ and $c_2 = a_5/(2a_4) \approx 1.623\cdot 10^{-4}\,\mathrm{K}^{-2}$. We derive equation (\ref{eq_mod}) by solving $\partial f/\partial \mathrm{RH}=0$ for RH. Condition (\ref{eq_mod}) of replacing RH triggers in roughly $1\%$ of our samples. It ensures that cloud cover does not increase when decreasing relative humidity in cases of low relative humidity and average temperature (see Fig~\ref{fig:phys_cons_adj}). Modifying the equation (\ref{best_eq}) in such a way does not deteriorate its performance on the DYAMOND data.
Fig~\ref{fig:phys_cons_adj}b illustrates how the modification ensures PC$_3$ in an average setting (in particular for $T = \overline{T}$). It would be difficult to apply a similar modification to the NN, which in our case violates PC$_3$ for $\mathrm{RH} > 0.95$. We can also directly identify another aspect of equation (\ref{best_eq}): the absence of a minimum value of relative humidity, below which cloud cover must always be zero (the \textit{critical relative humidity threshold}). 

Since $a_3 = (\partial I_1/ \partial T)\vert_{(\overline{\mathrm{RH}}, \overline{T})}$ is negative, cloud cover typically decreases with temperature for samples of the DYAMOND data set (see Fig~\ref{fig:analytic_eq}f)). However, $I_1$ does not ensure the PC$_6$ ($\partial {\cal C}/\partial T\leq 0$) constraint everywhere. For instance, in the hot limit $\lim_{T\rightarrow \infty} I_1(\mathrm{RH}, T)$, whether conditions are entirely cloudy or cloud-free depends upon relative humidity (in particular, whether $\mathrm{RH} > \overline{\mathrm{RH}}$).

The coefficient $a_4 = (\partial^2 I_1/\partial \mathrm{RH}^2)\vert_{(\overline{\mathrm{RH}}, \overline{T})}$ is precisely the curvature of $I_1$ w.r.t. $\mathrm{RH}$, causing the equation to flatten with decreasing $\mathrm{RH}$ (taking ($\ref{eq_mod}$) into account). It is consistent with the Sundqvist scheme that changes in relative humidity have a larger impact on cloud cover for larger relative humidity values. The final coefficient $a_5$ of $I_1$ is a third-order partial derivative of $I_1$ w.r.t. $T$ and $\mathrm{RH}$. More precisely,
\begin{equation*}
a_5 = \left. \left(\frac{\partial^3 I_1}{\partial T^2\partial RH}\right) \right\vert_{(\overline{\mathrm{RH}}, \overline{T})}.
\end{equation*}
The corresponding term becomes important whenever the temperature and relative humidity deviate strongly from their mean. In the upper or lower troposphere, where temperature conditions differ from the average tropospheric temperature, the $a_5$-term either further increases cloud cover in wet conditions (e.g., the tropical lower troposphere) or decreases it in dry conditions (e.g, in the upper troposphere or over the Sahara). The contribution of the $a_5$-term for selected vertical layers is illustrated in the second row of Fig~\ref{fig:geo_map}. When fit to the ERA5 data, the coefficients of the linear terms are found to be stable, while the emphasis on the non-linear terms is somewhat decreased; $a_4$ is $1.53$ and $a_5$ is $2.5$ times smaller. 

\subsection{Vertical Gradients in Relative Humidity and Stratocumulus Decks}
\label{sec:rh_grad_stratocumulus}

The second function $I_2(\partial_z \mathrm{RH})$ is a cubic polynomial of $\partial_z\mathrm{RH}$. Its magnitude is controlled by the coefficient $a_6$. If $a_6$ were 50\% smaller (which it is when fit to ERA5 data), it would decrease the absolute value of $I_2$ by 87.5\%. We introduce a prefactor of $1.5$ for $a_7$ so that $-a_7$ describes a local maximum of $I_2$ (found by solving $I_2'(\partial_zRH) = 0$). We will now focus on the reason for this distinct peak of $I_2 \approx 0.8$ at $\partial_zRH = -a_7$. \\
% The parameter $a_7$ indirectly defines a critical vertical relative humidity gradient; it is the minimal height over which $\mathrm{RH}$ can decrease by 10\% for the $I_2$-term to have a positive contribution to the cloud cover estimate. The specific value of $a_7 = 33\,$m (which is also stable on the ERA5 data set) ... \textcolor{red}{Physical meaning of these 33m? Literature?} \\
Removing the $I_2$-term, we find that the induced prediction error is largest, on average, in situations that are i) relatively dry (RH $ \approx 0.6$), ii) close to the surface ($z \approx 1000$m), iii) over water (land fraction $ \approx 0.1$), iv) characterized by an inversion ($\partial_zT \approx 0.01\,$K/m), and v) have small values of $\partial_z\mathrm{RH}$ ($\partial_z \mathrm{RH} \approx -2\,\text{km}^{-1} = -a_7$; compare also to the cloud cover peak in Fig~\ref{fig:analytic_eq}g). Using our cloud regimes of Sec~\ref{sec:cloud_regimes}, we find the average absolute error is largest in the stratus regime ($4\%$ cloud cover). Indeed, by plotting the globally averaged contributions of $I_1$, $I_2$ and $I_3$ on a vertical layer at about $1500$m altitude (Fig~\ref{fig:geo_map}), we find that $I_2$ is most active in regions with low-level inversions where marine stratocumulus clouds are abundant \cite{mauritsen2019}. From this, we can infer that the SFS NN has chosen $\partial_z\mathrm{RH}$ as a useful predictor to detect marine stratocumulus clouds and the symbolic regression algorithm has found a way to express this relationship mathematically. It is more informative than $\partial_zT$ (rank 10 in Sec~\ref{sec:feat_ranking}), which would measure the strength of an inversion more directly. Indeed, stratocumulus-topped boundary layers exhibit a sharp increase in temperature \textit{and} a sharp decrease in specific humidity between the cloud layer to the inversion layer. Studies by \citeA{nicholls1984dynamics} and \citeA{wood2012stratocumulus} reveal a notable temperature increase of approximately $5-6\,$K and a specific humidity decrease of about $4-5\,$g/kg. In ICON's grid with a vertical spacing of $\approx 300\,$m at an altitude of $1000 - 1500\,$m, the decrease in relative humidity would attain values of $\approx -2.5\,\text{km}^{-1}$. It is important to note that the vertical grid may not precisely separate the cloud layer from the inversion layer, making it reasonable to maximize the parameter $I_2$ at a relative humidity gradient of $\partial_z\mathrm{RH} = -2\,\text{km}^{-1}$. Vertical gradients of relative humidity below $-3,\text{km}^{-1}$ are extremely sporadic and confined to the lowest portion of the planetary boundary layer, where the vertical spacing between grid cells can get very small. In such cases, the attenuating effect of $I_2$ is unlikely to have significant physical causes. In contrast, vertical relative humidity gradients exceeding $1\,\text{km}^{-1}$ are common in the marine boundary layer due to evaporation and vertical mixing of moist air in the boundary layer. In this context, $I_2$ generally increases cloud cover which aligns with the fact that cloud cover is typically $5-15\%$ greater over the ocean compared to land \cite{rossow1999advances}. With the estimated values for $a_6$ and $a_7$, relative humidity would need to increase by $10\%$ over a height of $260\,$m to increase cloud cover by $10\%$.

\subsection{Understanding the Contribution of Cloud Condensates to Cloud Cover}

The third function $I_3(q_c, q_i)$ is always negative and decreases cloud cover where there is little cloud ice or water. It ensures that PC$_4$ and PC$_5$ are always satisfied. First of all, in condensate-free cells, $\epsilon$ serves to avoid division by zero while also decreasing cloud cover by $100\%$. Furthermore, the values of $a_8$ or $a_9$ indicate thresholds for cloud water/ice to cross to set $I_3$ closer to zero. When tuned to the ERA5 data set, the values for both $a_8$ and $a_9$ are roughly six times larger, making the equation less sensitive to cloud condensates. As larger values for cloud water are more common for cloud ice, we already expect $I_3$ to be more sensitive to cases when cloud ice actually does appear. By comparing the distributions of cloud ice/water at the storm-resolving scale, we provide a more rigorous derivation in \ref{app:sens} for why $a_9$ should indeed be smaller than $a_8$. A simple explanation is that we usually find ice clouds in the upper troposphere, where convection is associated with divergence, causing the clouds to spread out more.

% PC7 (/symbolic_regression/evaluate_schemes/analyzing_eqns/analyzing_eqns.ipynb)
Given that equation (\ref{best_eq}) is a continuous function, the continuity constraint PC$_7$ is only violated if and only if the cloud cover prediction is modified to be 0 in the condensate-free regime (by equation (\ref{eq:pc12})), and would be positive otherwise. The value of $\epsilon$ dictates how frequently the cloud cover prediction needs to be modified. In the limit $\epsilon \rightarrow 0$ we could remove the different treatment of the condensate-free case. In our data set, equation (\ref{best_eq}) yields a positive cloud cover prediction in $0.35\%$ of condensate-free samples. Thus, the continuity constraint PC$_7$ is almost always satisfied (in $99.65\%$ of our condensate-free samples).

\subsection{Ablation Study Confirms the Importance of Each Term}

\begin{figure}
\centering
\hspace*{-2.4cm}\includegraphics[width=1.3\textwidth]{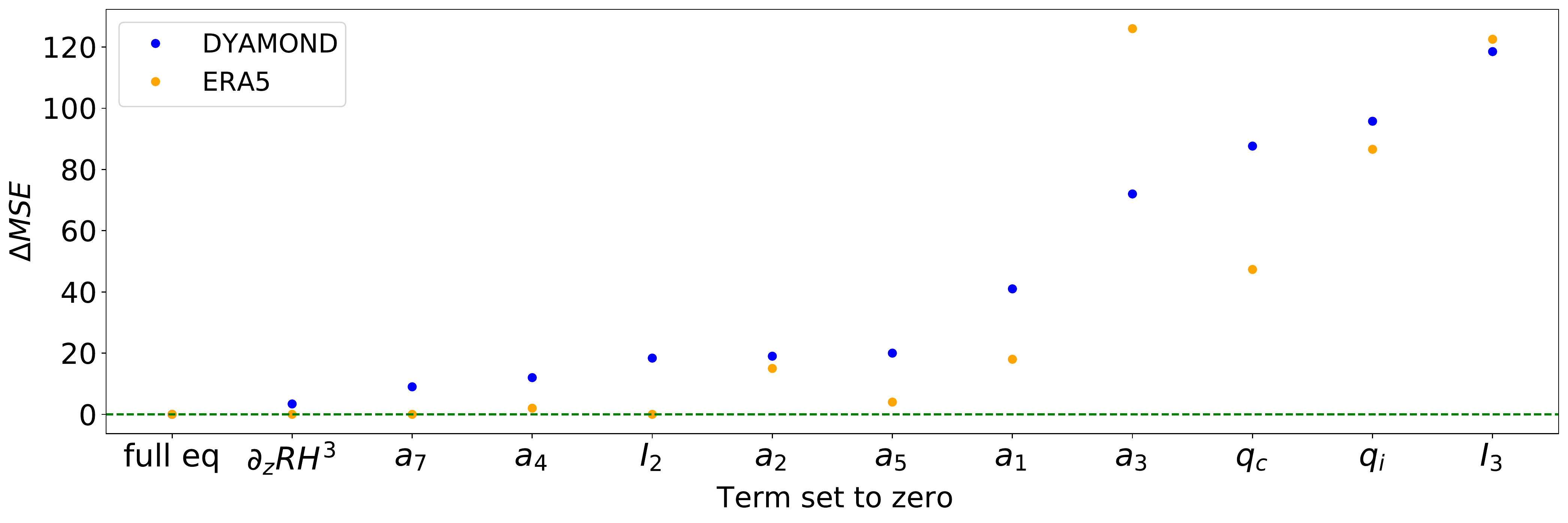}
\caption{Ablation study of equation (\ref{best_eq}) on the DYAMOND and ERA5 data sets. The removal of the function $I_1$ leads to a very large decrease of MSE (of $1300$/$763 \,(\%)^2$) on the DYAMOND/ERA5 data sets and is therefore not shown.}
\label{fig:ablation}
\end{figure}

To convince ourselves that all terms/parameters of equation (\ref{best_eq}) are indeed relevant to its skill, we examine the effects of their removal in an ablation study (Fig~\ref{fig:ablation}). We found that for the results to be meaningful, removing individual terms or parameters requires readjusting the remaining parameters; in a setting with fixed parameters the removal of multiple parameters often led to better outcomes than the removal of a single one of them. The optimizers (BFGS and Nelder-Mead) used to retune the remaining parameters show different success depending on whether the removal of terms is applied to the equation formulated in terms of normalized or physical features (the latter being equation (\ref{best_eq})). Therefore, each term is removed in both formulations, and the better result is chosen each time. To ensure robustness of the results, this ablation study is repeated for $10$ different seeds on subsets with $10^6$ data samples. \\
We find that the removal of any individual term in equation (\ref{best_eq}) would result in a noticeable reduction in performance on the DYAMOND data ($\Delta MSE \geq 3.4 \,(\%)^2$ in absolute and $(MSE_{abl}-MSE_{full})/MSE_{abl} \geq 3.2\%$ in relative terms). Even though Fig~\ref{fig:analytic_eq}g) suggests a cubic dependence of cloud cover on $\partial_z\mathrm{RH}$, it is the least important term to include according to Fig~\ref{fig:ablation}. Applied to the ERA5 data, we can even dispense with the entire $I_2$ term. Furthermore, we find that the quadratic dependence on RH can be largely compensated by the linear terms. The most important terms to include are those with cloud ice/water and the linear dependence on temperature. 
Coinciding with the SFS NN feature sequences in Sec~\ref{sec:feat_ranking}, cloud ice ($\Delta MSE = 96/102 \,(\%)^2$) is more important to take into account than cloud water ($\Delta MSE = 88/63 \,(\%)^2$), especially for the ERA5 data set in which cloud ice is more abundant (see Fig~\ref{fig:era5}). More generally, out of the functions $I_1$, $I_2$, $I_3$ we find $I_1(\mathrm{RH}, T)$ to be most relevant ($\Delta MSE = 1300/763 \,(\%)^2$), followed by $I_3(q_c, q_i)$ ($\Delta MSE = 119/123 \,(\%)^2$) and lastly $I_2(\partial_z\mathrm{RH})$ ($\Delta MSE = 18/0 \,(\%)^2$), once again matching the order of features that the SFS NNs had chosen.

\section{Conclusion}
%% Summary, highlights, contributions
In this study, we derived data-driven cloud cover parameterizations from coarse-grained global storm-resolving simulation (DYAMOND) output. We systematically populated a performance $\times$ complexity plane with interpretable traditional parameterizations and regression fits on one side and high-performing neural networks on the other. Modern symbolic regression libraries (PySR, GPGOMEA) allow us to discover interpretable equations that diagnose cloud cover with excellent accuracy ($R^2 > 0.9$). From these equations, we propose a new analytical scheme for cloud cover (found with PySR) that balances accuracy ($R^2 = 0.94$) and simplicity (10 free parameters in the physical formulation). This analytical scheme satisfies six out of seven physical constraints (although the continuity constraint is violated in $0.35\%$ of our condensate-free samples), providing the crucial third criterion for its selection. In a first evaluation, the (5-feature) analytical scheme was on par with the 6-feature NN in terms of reproducing cloud cover distributions (Hellinger distances $< 0.05$) in condensate-rich cloud regimes, yet underestimating cloud cover more strongly in condensate-poor regimes. While discovering distinct equations in each cloud regime can improve the Hellinger distances, both the overall complexity and mean squared error of a combined piecewise equation increase. This supports choosing a single continuous analytical scheme that generalizes across cloud regimes. When applied to higher resolutions than their training data we find that the cloud cover schemes further improve their performance. This finding opens up possibilities for leveraging their predictive capabilities in domains with increased resolution requirements.

In addition to its interpretability, flexibility and efficiency, another major advantage of our best analytical scheme is its ability to adapt to a different data set (in our case, the ERA5 reanalysis product) after learning from only a few of the ERA5 samples in a transfer learning experiment. Due to the small amount of free parameters and the initial good fit on the DYAMOND data, our new analytical scheme outperformed all other Pareto-optimal models. We found that as the number of samples in the transfer learning sets increases, the models converged to the same performance rank on the ERA5 data as on the DYAMOND data, indicating strong similarities in the nature of the two data sets that could make which data set serves as the training set irrelevant. In an ablation study, we found that further reducing the number of free parameters in the analytical scheme would be inadvisable; all terms/parameters are relevant to its performance on the DYAMOND data. Key terms include a polynomial dependence on relative humidity and temperature, and a nonlinear dependence on cloud ice and water.

% Frame it in the context of climate modeling/understanding
% dzRH and marine stratomuli
Our sequential feature selection approach with NNs revealed an objectively good subset of features for an unknown nonlinear function: relative humidity, cloud ice, cloud water, temperature and the vertical derivative of relative humidity (most likely linked to the vertical variability of cloud cover within a grid cell). While the first four features are well-known predictors for cloud cover, PySR also learned to incorporate $\partial_zRH$ in its equation. This additional dependence allows it to detect thin marine stratocumulus clouds, which are difficult, if not impossible to infer from exclusively local variables. These clouds are notoriously underestimated in the vertically coarse climate models \cite{nam2012too}. In ICON this issue is somewhat attenuated by multiplying, and thus increasing relative humidity in maritime regions by a factor depending on the strength of the low-level inversion \cite{mauritsen2019}. Using symbolic regression, we thus found an alternative, arguably less crude approach, which could help mitigate this long-standing bias in an automated fashion. However, we need to emphasize that in particular shallow convection is not yet properly resolved on kilometer-scale resolutions. Therefore, shallow clouds such as stratocumulus clouds are still distorted in the storm-resolving simulations we use as the source of our training data \cite{stevens2020added}. To properly capture shallow clouds it could be advisable to further increase the resolution of the high-resolution model, training on coarse-grained output from targeted large-eddy simulations \cite{stevens2005evaluation} or observations.

% Mention online coupling. Explicitly mention why it's left for future work.
A crucial next step will be to test the cloud cover schemes when coupled to Earth system models, including ICON. We decided to leave this step for future work for several reasons. First, our focus was on the equation discovery methodology and the analysis of the discovered equation. Second, our goal was to derive a cloud cover scheme that is climate model-independent. Designing a scheme according to its online performance within a specific climate model decreases the likelihood of inter-model compatibility as the scheme has to compensate the climate model's parameterizations' individual biases. For instance, in ICON, the other parameterizations would most likely need to be re-calibrated to adjust for current compensating biases, such as clouds being `too few and too bright' \cite{crueger2018icon}. Third, the metrics used to validate a coupled model remain an active research area, and at this point, it is unclear which targets must be met to accept a new ML-based parameterization. That being said, the superior transferability of our analytical scheme to the ERA5 reanalysis data not only suggests its applicability to observational data sets, but also that it may be transferable to other Earth system models.

% Limitations
In addition to inadequacies in our training data (see above), which somewhat exacerbate the physical interpretation of the derived analytical equations, our current approach has some limitations. Symbolic regression libraries are limited in discovering equations with a large number of features. In many cases, five features are insufficient to uncover a useful data-driven equation, requiring a reduction of the feature space's dimensionality. To measure model complexity, we used the number of free parameters, disregarding the number of features and operators. Although the number of operators in our study was roughly equivalent to the number of parameters, this may not hold in more general applications and the complexity of individual operators would need to be specified (as in \ref{app:pysr}).

% Compare our approach to another approach (Ross)/Follow-up tasks
Our approach differs from similar methods used to discover equations for ocean subgrid closures \cite{ross2023benchmarking, zanna2020data} because we included nonlinear dependencies without assuming additive separability, instead fitting the entire equation non-iteratively. By simply allowing for division as an operator in our symbolic regression method, we found rational nonlinearities in the equation whose detection would already require modifications such as \citeA{kaheman2020sindy} to conventional sparse regression approaches. Despite our efforts, the equation we found is still not as accurate as an NN with equivalent features in the cirrus-like regime (the Hellinger distance between the analytical scheme and the DYAMOND cloud cover distribution was more than twice as large as for the NN). Comparing the partial dependence plots of the equation with those of the NN could provide insights and define strategies to further extend and improve the equation, while reducing the computational cost of the discovery. There are various methods available for utilizing NNs in symbolic regression for more than just feature selection, one of which is AIFeynman \cite{udrescu2020ai}. While AIFeynman is based on the questionable assumption that the gradient of an NN provides useful information, a direct prediction of the equation using recurrent neural networks presents a promising avenue for improved symbolic regression \cite{petersen2021deep,tenachi2023deep}.

% Nonetheless, our study demonstrated the potential of data-driven equation discovery for improving cloud cover parameterizations in Earth system models. However, the overall framework is not only applicable to climate research, but can also be used to aid several other scientific disciplines that seek to derive interpretable equations directly from data.

Nonetheless, our simple cloud cover equation already achieves high performance. Our study thus underscores that symbolic regression can complement deep learning by deriving interpretable equations directly from data, suggesting untapped potential in other areas of Earth system science and beyond. 

\newpage

\appendix
\section{Global Maps of $I_1$, $I_2$, $I_3$} \label{app:global_maps}
\begin{figure}
\centering
\hspace*{-2.6cm}\includegraphics[width=1.4\textwidth]{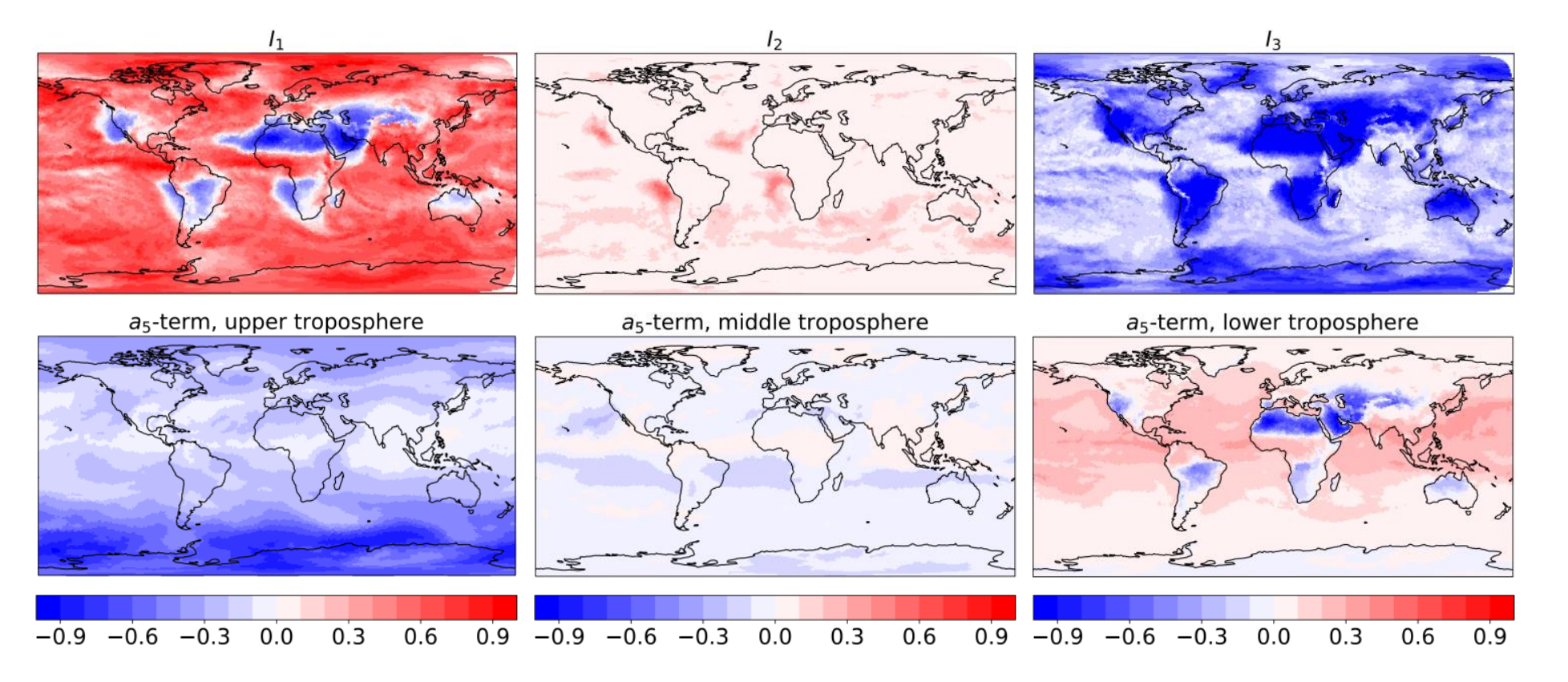}
\caption{The first row shows maps of $I_1(\mathrm{RH}, T)$, $I_2(\partial_zRH)$ and $I_3(q_c, q_i)$ on a vertical layer with an average height of 1490m. In the second row we zoom in on the contribution of the term in $I_1$ corresponding to the $a_5$-coefficient on three different height levels (roughly at $11\,$km, $4\,$km, $320\,$m). All plots are averaged over 10 days (11 Aug\textendash{}20 Aug, 2016). The data source is the coarse-grained three-hourly DYAMOND data.}
\label{fig:geo_map}
\end{figure}

In this section, we plot average function values for the three terms $I_1$, $I_2$, and $I_3$ of equation (\ref{best_eq}). We focus on the vertical layer roughly corresponding to an altitude of $1500\,$m to analyze if one of the terms would detect thin marine stratocumulus clouds. Due to their small vertical extent, these clouds are difficult to pick up on in coarse climate models, which constitutes a well-known bias. To compensate for this bias, the current cloud cover scheme of ICON has been modified so that relative humidity is artificially increased in low-level inversions over the ocean \cite{mauritsen2019}. \\
Analyzing Fig~\ref{fig:geo_map}, we find that the regions of high $I_2$-values correspond with regions typical for low-level inversions and low-cloud fraction \cite{mauritsen2019, muhlbauer2014climatology}. These $I_2$-values compensate partially negative $I_1$- and $I_3$-values in low-cloud regions of the Northeast Pacific, Southeast Pacific, Northeast Atlantic, and the Southeast Atlantic. The $I_3$-term decreases cloud cover over land and is mostly inactive over the oceans due to the abundancy of cloud water. The $I_1$-term is particularly small in the dry and hot regions of the Sahara and the Rub' al Khali desert and largest over the cold poles. The $a_5$-term is the only term in $I_1$ that cannot be explained as a linear or a curvature term. In the upper troposphere, the term is negative due to relatively cold and dry conditions. In August, temperatures are coldest in the southern hemisphere, so the term has a strong negative effect, especially over the South Pole. In the middle troposphere, temperatures are near the average of $257\,$K, weakening the term overall. Negative patches in the subtropics are due to the dry descending branches of the Hadley cell. The lower troposphere is relatively warm, especially in the tropics, resulting in a large positive $a_5$-term under humid conditions, and a negative term under dry conditions.

\section{The Sensitivity of Cloud Cover to Cloud Water and Ice \label{app:sens}}

\begin{figure}
\centering
\hspace*{-1.2cm}\includegraphics[width=1.1\textwidth]{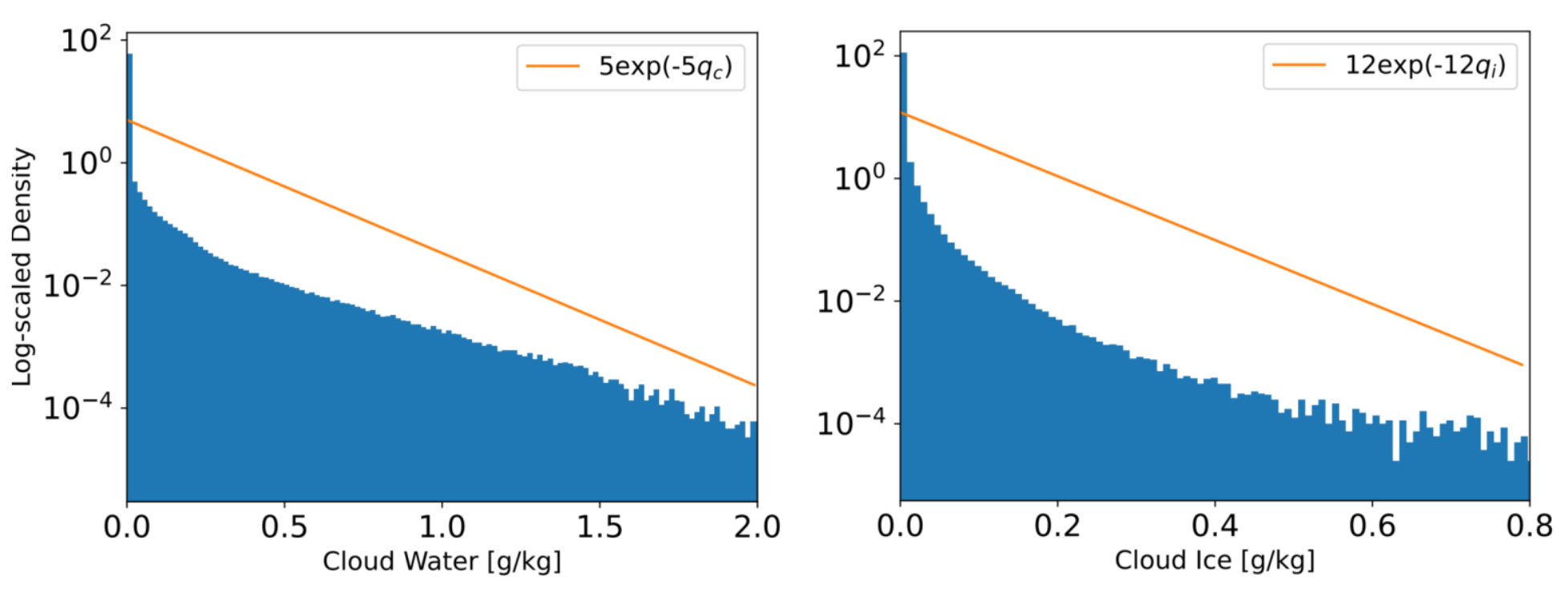}
\caption{The distributions of cloud water and cloud ice on storm-resolving scales ($2.5\,$km DYAMOND Winter data). For positive values we approximate these distributions very loosely with exponential distributions.}
\label{fig:cloud_cond_dya}
\end{figure}

In Equation (\ref{best_eq}), cloud cover is more sensitive to cloud ice than cloud water. In this section, we show that we can explain this difference in sensitivity from the storm-scale distributions of cloud water and ice alone (Fig~\ref{fig:cloud_cond_dya}). 
On storm-resolving scales, a grid cell is fully cloudy if cloud condensates $q_t$ exceed a small threshold $a > 0$. Otherwise it is set to be non-cloudy. We can thus express the expected cloud cover as the probability of $q_t$ exceeding the threshold $a$

\begin{equation}
\label{eq:appC}
    \mathbb{E}[\mathcal{C}] = \mathbb{P}[q_t > a] = \int_a^{\infty} f_{q_t}(q_t) dq_t, 
\end{equation}
where $f_{x}$ is the probability density function of some variable $x$. As we can express cloud condensates as a sum of cloud water $q_c$ and cloud ice $q_i$, we can also derive $f_{q_t}$ from $f_{q_c}$ and $f_{q_i}$ by fixing $q_t$ and integrating over all potential values for $q_c$

\begin{equation}
\label{eq:appconv}
    f_{q_t}(q_t) = \int_0^{q_t} f_{q_c}(z) f_{q_i}(q_t - z) dz.
\end{equation}
In the following, we introduce the subscript $s$ as a placeholder for either liquid or ice. According to Fig~\ref{fig:cloud_cond_dya}, the storm-resolving cloud ice/water distributions feature distinct peaks at $q_s = 0$, which can be modeled by weighted dirac-delta distributions. For $q_s > 0$, we can approximate $f_{q_c}$ and $f_{q_i}$ with exponential distributions. After normalizing the distributions so that their integrals over $q_s \geq 0$ yield $1$ we arrive at
\begin{equation*}
    f_{q_s}(q_s) = (\lambda_s \exp(-\lambda_s q_s) + w_s\delta(q_s))/(w_s/2 + 1).
\end{equation*}
By rephrasing $w_s$ in terms of $\lambda_s$ and $\mu_s$, the mean of $f_{q_s}$, we get
\begin{equation}
\label{eq:appfqs}
    f_{q_s}(q_s) = \lambda_s\mu_s(\lambda_s\exp(-\lambda_s q_s) + (-2 + 2/(\lambda_s\mu_s))\delta(q_s)).
\end{equation}
By plugging in the expressions (\ref{eq:appfqs}) and (\ref{eq:appconv}) into equation (\ref{eq:appC}) and letting $a \rightarrow 0^{+}$ we find the expected cloud cover to be a function of the shape parameters $\lambda_s$ and the means $\mu_s$ for cloud water and ice
\begin{equation}
\label{eq:appcf}
    \mathbb{E}[\mathcal{C}] = -3\lambda_{i}\lambda_{c}\mu_{i}\mu_{c} + 2\lambda_{i}\mu_{i} + 2\lambda_{c}\mu_{c}.
\end{equation}
We can relate this expression to $a_8$ and $a_9$ by expanding $I_3$ to first order around the origin
\begin{equation}
\label{eq:apptay}
    I_3(q_c, q_i) \approx -1/\epsilon + q_c/(a_8 \epsilon^2) + q_i/(a_9 \epsilon^2) - q_c q_i/(a_8 a_9 \epsilon^3).
\end{equation}
By comparing (\ref{eq:appcf}) and (\ref{eq:apptay}) we arrive at the following analogy for $q_s \approx \mu_s$:
\begin{equation*}
    2\lambda_l \approx 1/(a_8\epsilon^2) \text{ and } 2\lambda_i \approx 1/(a_9\epsilon^2).
\end{equation*}
We conclude that the larger the shape parameter, i.e., the faster the distribution tends to zero, the smaller we expect the associated parameter to be. Based on Fig~\ref{fig:cloud_cond_dya} we have $\lambda_i > \lambda_c$, which explains why $a_9$ is smaller than $a_8$. In other words, why $I_3$ is more sensitive to cloud ice than cloud water.
\section{PySR Settings \label{app:pysr}}
First of all, we do not restrict the number of iterations, and instead restrict the runtime of the algorithm to $\approx\,8$ hours. We choose a large set of operators $O$ to allow for various different functional forms (while leaving out non-continuous operators). To aid readability we show the operators applied to some $\left(x,y\right) \in \mathbb{R}^2$ which we denote by superscripts. To account for the different complexity of the operators, we split $O$ into four distinct subsets 
\begin{align*}
O_1^{(x,y)} &= \{x \cdot y, x+y, x-y, -x\}\\
O_2^{(x,y)} &= \{x/y, \vert x \vert, \sqrt{x}, x^3, \max(0, x)\} \\
O_3^{(x,y)} &= \{\exp(x), \ln(x), \sin(x), \cos(x), \tan(x), \sinh(x), \cosh(x), \tanh(x)\} \\
O_4^{(x,y)} &= \{x^y, \Gamma(x), \text{erf}(x), \arcsin(x), \arccos(x), \arctan(x), \text{arsinh}(x), \text{arcosh}(x), \text{artanh}(x)\}
\end{align*}
of increasing complexity. The operators in $O_2$/$O_3$/$O_4$ are set to be $2/3/9$ times as complex as those in $O_1$. In this manner, for instance $x^3$ and $(x\cdot x)\cdot x$ have the same complexity. Furthermore, we assign a relatively low complexity to the operators in $O_3$ as they are very common and have well-behaved derivatives. With the factor of $9$, we strongly discourage operators in $O_4$. We expect that for every occurrence of a variable in a candidate equation it will also need to be scaled by a certain factor. We do not want to discourage the use of such constant factors or the use of variables themselves and leave the complexity of constants and variables at their default complexity of one.

We obtain the best results when setting the complexity of the operators in $O_1$ to $3$ and training the PySR scheme on $5000$ random samples. Other parameters include the population size (set to $20$) and the maximum complexity of the equations that we initially set to $200$ and reduced to $90$ in later runs. 

\section{Selected Symbolic Regression Fits}
\label{app:eqns}
This section lists all equations found with the symbolic regression libraries GP-GOMEA or PySR that are included in Fig~\ref{fig:pareto_plot}, ranked in increasing MSE order. In brackets we provide the MSE/number of parameters. We list the equations according to their MSE. The equations that lie on the Pareto frontier are highlighted in bold:
\begin{align*}
%eq4
& \text{1) PySR }[103.95/11]: \\
& f(\mathrm{RH}, T, \partial_z\mathrm{RH}, q_c, q_i) = \boldsymbol{203\mathrm{RH}^2 + (0.06588\mathrm{RH} - 0.03969)T^2 - 33.87\mathrm{RH}T + 4224.6\mathrm{RH}} \\ 
& \boldsymbol{ + 18.9586T - 2202.6 + (2\cdot 10^{10} \partial_z\mathrm{RH} + 6\cdot 10^7)(\partial_z\mathrm{RH})^2 - 1/(8641q_c + 32544q_i + 0.0106)} \\
%eqC
& \text{2) PySR }[104.26/19]: \\
& f(\mathrm{RH}, T, \partial_z\mathrm{RH}, q_c, q_i) =  (1.0364\mathrm{RH} - 0.6782)(0.0581T - 16.1884)(-44639.6\partial_z\mathrm{RH} + 1.1483T - 262.16) \\
&+ 171.963\mathrm{RH} - 1.4705T + 158.433(\mathrm{RH} - 0.60251)^2 + (\partial_z\mathrm{RH})^2(2 \cdot 10^{11}q_c - 8 \cdot 10^7 \mathrm{RH} + 7\cdot 10^7) + 316.157 \\
&+ 93319q_i - 1/(12108q_c + 39564q_i + 0.0111) \\
%eq3
& \text{3) PySR }[106.52/12]:\\
& f(\mathrm{RH}, T, \partial_z\mathrm{RH}, q_c, q_i) = (57.2079\mathrm{RH} - 34.4685)(3.0985\mathrm{RH} + 73.1646(0.0039T - 1)^2 - 1.8669) + 123.175\mathrm{RH} \\
&- 1.4091T + 1.5\cdot 10^7(\partial_z\mathrm{RH})^2(10619q_c - 4.9155\mathrm{RH} + 4.7178) + 333.1 - 1/(10367q_c + 35939q_i + 0.0111) \\
%eq2
& \text{4) PySR }[106.95/11]:\\
& f(\mathrm{RH}, T, \partial_z\mathrm{RH}, q_c, q_i) = 19.3885(3.0076\mathrm{RH} - 1.8121)(3.2825\mathrm{RH} + 73.1646(0.0039T - 1)^2 - 1.9777) \\
& + 118.59\mathrm{RH} - 1.423T + 1.5\cdot 10^7(3.0125 - 1.0129\mathrm{RH})(\partial_z\mathrm{RH})^2 + 339.2 - 1/(9325q_c + 34335q_i + 0.0109) \\
%eq1
& \text{5) PySR }[106.99/10]:\\
& f(\mathrm{RH}, T, \partial_z\mathrm{RH}, q_c, q_i) = \boldsymbol{(58.189\mathrm{RH} - 35.0596)(3.3481\mathrm{RH} + 73.1646(0.0039T - 1)^2 - 2.0172)} \\
&\boldsymbol{+ 116.873\mathrm{RH} - 1.4211T + 3.6\cdot 10^7(\partial_z\mathrm{RH})^2 + 339.9 - 1/(9237q_c + 34136q_i + 0.0109)} \\
%eq5
& \text{6) PySR }[111.76/15]: \\
& f(\mathrm{RH}, T, \partial_z\mathrm{RH}, q_c, q_i) = (3.2665\mathrm{RH} - 2.9617)(0.0435T - 9.0274)(16073.2\partial_z\mathrm{RH} + 0.3013T - 68.4342) \\
&97.5754\mathrm{RH} - 0.6556T + 175 + 123823q_i - 1/(9853q_c + 36782q_i + 0.0112) \\
%eq1
& \text{7) GP-GOMEA }[121.89/13]: \\ 
& f(\mathrm{\mathrm{RH}}, T, q_c, q_i) = 8.459\exp(2.559\mathrm{RH}) - 33.222\sin(0.038T + 109.878) + 24.184 \\ & -\sin(3.767\sqrt{\vert 98709q_i - 0.334\vert})/(30046q_i + 5628q_c + 0.01) \\
%eq2
& \text{8) GP-GOMEA }[136.64/11]: \\
& f(\mathrm{\mathrm{RH}}, T, q_c, q_i) = (8.65\mathrm{RH} - 0.22T - 93.14)\sqrt{\vert 0.62T - 414.23\vert} + 2368 - 1/(28661q_i + 4837q_c + 0.01) \\
%eq3
& \text{9) GP-GOMEA }[159.80/9]: \\
& f(\mathrm{\mathrm{RH}}, q_c, q_i) = \boldsymbol{0.009 e^{8.725\mathrm{RH}} + 12.795\log(229004q_i + 0.774(e^{11357q_c} - 1)) - 178246q_c + 66}\\
%eq4
& \text{10) GP-GOMEA }[161.45/12]: \\
& f(\mathrm{\mathrm{RH}}, T, q_c, q_i) = (0.028e^{6.253\mathrm{RH}} + 5\mathrm{RH} - 0.076T + 4)/(183894q_i + 0.73e^{6565q_c - 91207q_i} - 0.62) + 92.3
\end{align*}

Note that the assessed number of parameters is based on a simplified form of the equations in terms of its normalized variables. The amount of parameters in a given equation is at least equal to the assessed number of parameters minus one (accounting for the zero in the condensate-free setting).

% %%%%%%%%%%%%%%%%%%%%%%%%%%%%%%%%%%%%%%%%%%%%%%%%%%%%%%%%%%%%%%%%
% %
% % Optional Glossary, Notation or Acronym section goes here:
% %
% %%%%%%%%%%%%%%
% % Glossary is only allowed in Reviews of Geophysics
% %  \begin{glossary}
% %  \term{Term}
% %   Term Definition here
% %  \term{Term}
% %   Term Definition here
% %  \term{Term}
% %   Term Definition here
% %  \end{glossary}

% %
% %%%%%%%%%%%%%%
% % Acronyms
% %   \begin{acronyms}
% %   \acro{Acronym}
% %   Definition here
% %   \acro{EMOS}
% %   Ensemble model output statistics
% %   \acro{ECMWF}
% %   Centre for Medium-Range Weather Forecasts
% %   \end{acronyms}

% %
% %%%%%%%%%%%%%%
% % Notation
% %   \begin{notation}
% %   \notation{$a+b$} Notation Definition here
% %   \notation{$e=mc^2$}
% %   Equation in German-born physicist Albert Einstein's theory of special
% %  relativity that showed that the increased relativistic mass ($m$) of a
% %  body comes from the energy of motion of the body—that is, its kinetic
% %  energy ($E$)—divided by the speed of light squared ($c^2$).
% %   \end{notation}

\vskip12pt\noindent{\bf
Open Research\vrule depth 6pt
width0pt\relax}\\*\noindent\ignorespaces
% For DYAMOND see https://www.esiwace.eu/services/dyamond-initiative/acknowledging
The cloud cover schemes and analysis code are preserved \cite{sweqdisc}. DYAMOND data management was provided by the German Climate Computing Center (DKRZ) and supported through the projects ESiWACE and ESiWACE2. The coarse-grained model output used to train and evaluate the neural networks amounts to several TB and can be reconstructed with the scripts provided in the GitHub repository. 

% It is important to cite individual data sets in this section and, and they must be included in your bibliography. Please use the type field in your bibtex file to specify the type of data cited. Some options include data set, Software, Collection, ComputationalNotebook. Ex: 
% \\
% \begin{verbatim}

% @misc{https://doi.org/10.7283/633e-1497,
%   doi = {10.7283/633E-1497},
%   url = {https://www.unavco.org/data/doi/10.7283/633E-1497},
%   author = {de Zeeuw-van Dalfsen, Elske and Sleeman, Reinoud},
%   title = {KNMI Dutch Antilles GPS Network - SAB1-St_Johns_Saba_NA P.S.},
%   publisher = {UNAVCO, Inc.},
%   year = {2019},
%   type = {data set}
% }

% \end{verbatim}

% For physical samples, use the IGSN persistent identifier, see the International Geo Sample Numbers section:
% \url{https://www.agu.org/Publish-with-AGU/Publish/Author-Resources/Data-and-Software-for-Authors#IGSN}
% %%%%%%%%%%%%%%%%%%%%%%%%%%%%%%%%%%%%%%%%%%%%%%%

\acknowledgments
% For DYAMOND see https://www.esiwace.eu/services/dyamond-initiative/acknowledging
Funding for this study was provided by the European Research Council (ERC) Synergy Grant “Understanding and Modelling the Earth System with Machine Learning (USMILE)” under the Horizon 2020 research and innovation programme (Grant agreement No. 855187). Beucler acknowledges funding from the Columbia University sub-award 1 (PG010560-01). Gentine acknowledges funding from the NSF Science and Technology Center, Center for Learning the Earth with Artificial Intelligence and Physics (LEAP) (Award 2019625). This manuscript contains modified Copernicus Climate Change Service Information (2023) with the following datasets being retrieved from the Climate Data Store:
ERA5, ERA5.1 (neither the European Commission nor ECMWF is responsible for
any use that may be made of the Copernicus Information or Data it contains). The projects ESiWACE and ESiWACE2 have received funding from the European Union’s Horizon 2020 research and innovation programme under grant agreements No 675191 and 823988. This work used resources of the Deutsches Klimarechenzentrum (DKRZ) granted by its Scientific Steering Committee (WLA) under project IDs bk1040, bb1153 and bd1179.

% %% ------------------------------------------------------------------------ %%
% %% References and Citations

% %%%%%%%%%%%%%%%%%%%%%%%%%%%%%%%%%%%%%%%%%%%%%%%
% %
\bibliography{bibfile}
% %
% % don't specify bibliographystyle

% % In the References section, cite the data/software described in the Availability Statement (this includes primary and processed data used for your research). For details on data/software citation as well as examples, see the Data & Software Citation section of the Data & Software for Authors guidance
% % https://www.agu.org/Publish-with-AGU/Publish/Author-Resources/Data-and-Software-for-Authors#citation

% %%%%%%%%%%%%%%%%%%%%%%%%%%%%%%%%%%%%%%%%%%%%%%%

% %\bibliography{enter your bibtex bibliography filename here}

% %Reference citation instructions and examples:
% %
% % Please use ONLY \cite and \citeA for reference citations.
% % \cite for parenthetical references
% % ...as shown in recent studies (Simpson et al., 2019)
% % \citeA for in-text citations
% % ...Simpson et al. (2019) have shown...
% %
% %
% %...as shown by \citeA{jskilby}.
% %...as shown by \citeA{lewin76}, \citeA{carson86}, \citeA{bartoldy02}, and \citeA{rinaldi03}.
% %...has been shown \cite{jskilbye}.
% %...has been shown \cite{lewin76,carson86,bartoldy02,rinaldi03}.
% %... \cite <i.e.>[]{lewin76,carson86,bartoldy02,rinaldi03}.
% %...has been shown by \cite <e.g.,>[and others]{lewin76}.
% %
% % apacite uses < > for prenotes and [ ] for postnotes
% % DO NOT use other cite commands (e.g., \citet, \citep, \citeyear, \citealp, etc.).
% % \nocite is okay to use to add references from your Supporting Information
% %

\end{document}

%% ------------------------------------------------------------------------ %%
%
%  TITLE
%
%% ------------------------------------------------------------------------ %%

%\includegraphics{agu_pubart-white_reduced.eps}

%\graphicspath{{figures/}}

\title{Supporting Information for ``Data-Driven Equation Discovery of a Cloud Cover Parameterization"}
%
% e.g., \title{Supporting Information for "Terrestrial ring current:
% Origin, formation, and decay $\alpha\beta\Gamma\Delta$"}
%
%DOI: 10.1002/%insert paper number here%

%% ------------------------------------------------------------------------ %%
%
%  AUTHORS AND AFFILIATIONS
%
%% ------------------------------------------------------------------------ %%

% List authors by first name or initial followed by last name and
% separated by commas. Use \affil{} to number affiliations, and
% \thanks{} for author notes.
% Additional author notes should be indicated with \thanks{} (for
% example, for current addresses).

\authors{Arthur Grundner\affil{1,2}, Tom Beucler\affil{3}, Pierre Gentine\affil{2}, and Veronika Eyring\affil{1,4}}

\affiliation{1}{Deutsches Zentrum für Luft- und Raumfahrt e.V. (DLR), Institut für Physik der Atmosphäre, Oberpfaffenhofen, Germany}
\affiliation{2}{Center for Learning the Earth with Artificial Intelligence And Physics (LEAP), Columbia University, New York, NY, USA}
\affiliation{3}{Institute of Earth Surface Dynamics, University of Lausanne, Lausanne, Switzerland}
\affiliation{4}{University of Bremen, Institute of Environmental Physics (IUP), Bremen, Germany}

\noindent\textbf{Contents}
%%%Remove or add items as needed%%%
\begin{enumerate}
% \item Text S1 to S3
% \item Tables S1 to S2
\item Text S1 
\item Figures S1 to S3
\end{enumerate}

\clearpage

\begin{figure}
    \centering
    \hspace*{-2cm}\includegraphics[scale=0.54]{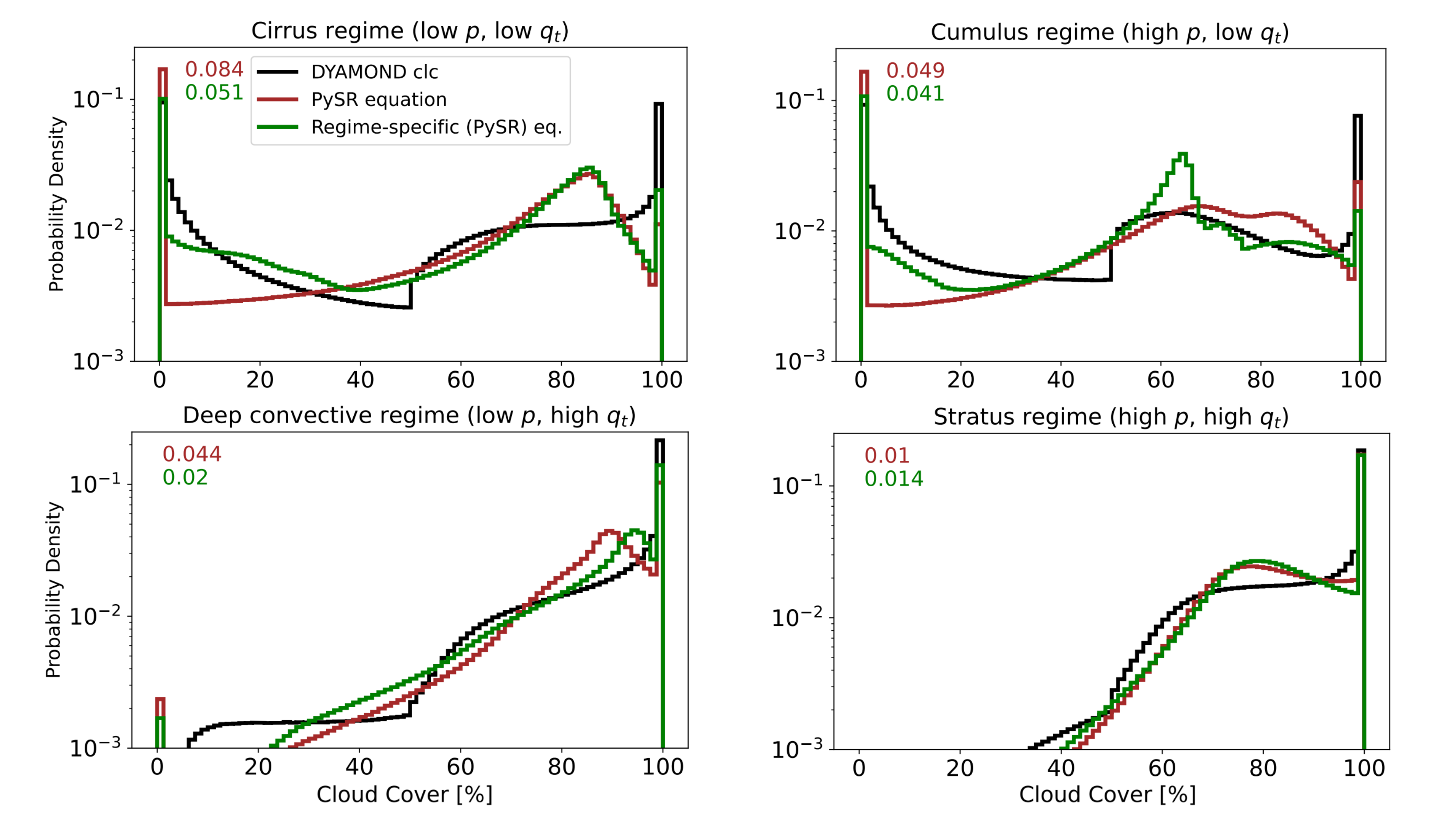}
    \caption{Predicted cloud cover distributions of the selected PySR equation of Sec~5.2 and Sec~6 (discovered on the entire coarse-grained DYAMOND data set) and of regime-specific equations found with PySR (for the functional form see above). Each panel corresponds to a distinct cloud regime (cf. Sec~5.2). The numbers in the upper left indicate the Hellinger distance between the predicted and the actual cloud cover distributions for each model and cloud regime.}
    \label{fig:pysr_reg_specific}
\end{figure}

\noindent\textbf{Text S1.}
Here we provide the functional form of the regime-specific equations of Fig~\ref{fig:pysr_reg_specific}:
\begin{align}
    f_{cirrus}(q_i, \mathrm{RH}, T) &= \frac{(3.008\mathrm{RH} - 0.03327T + 8.245)(3.008\mathrm{RH} + 3733000q_i - 1.558)}{98710q_i + 0.06077} \\
    f_{cumulus}(q_i, q_c, \mathrm{RH}) &= 126.3\mathrm{RH}-1871000q_c-8.046 - \frac{5.215}{17550q_c+98710q_i+0.05212}\\
    f_{deep\,conv.}(\mathrm{RH}, T, \partial_z\mathrm{RH}) &= -34860\,\partial_z\mathrm{RH} - 1.34T  + 387\\
    &+ 120.6(\mathrm{RH}-0.6)\left((0.033T - 8.55)(27.2(\mathrm{RH}-0.6)^3-0.6) + 1.4 \right)\\
    f_{stratus}(\mathrm{RH}, \partial_z\mathrm{RH}) &=3744\,\partial_z\mathrm{RH} + 39310000\,\partial_z\mathrm{RH}^2 + 7.221e^{3\,\mathrm{RH}} - 38.64,
\end{align}
where the features have been normalized over the training set. The total number of free trainable parameters is $33$ ($8 + 6 + 11 + 5$ for the regime-specific equations above + 2 for the switch between cloud regimes + 1 for the condensate-free regime).

\begin{figure}
    \centering
    \hspace*{-3.0cm}\includegraphics[scale=0.6]{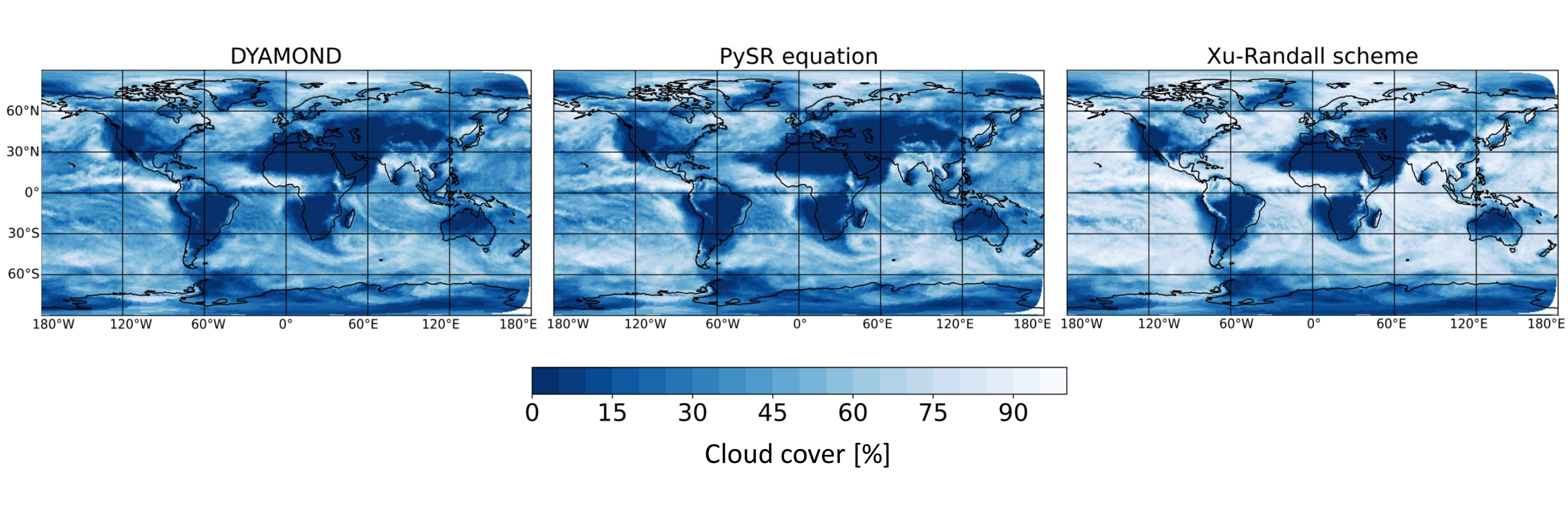}
    \caption{Averaged cloud cover from 11 Aug to 20 Aug, 2016 at an altitude of $\approx\,$1500 m. The data source is the coarse-grained three-hourly DYAMOND data.}
    \label{fig:pysr_1500m}
\end{figure}

\begin{figure}
    \centering
    \hspace*{-3.0cm}\includegraphics[scale=0.6]{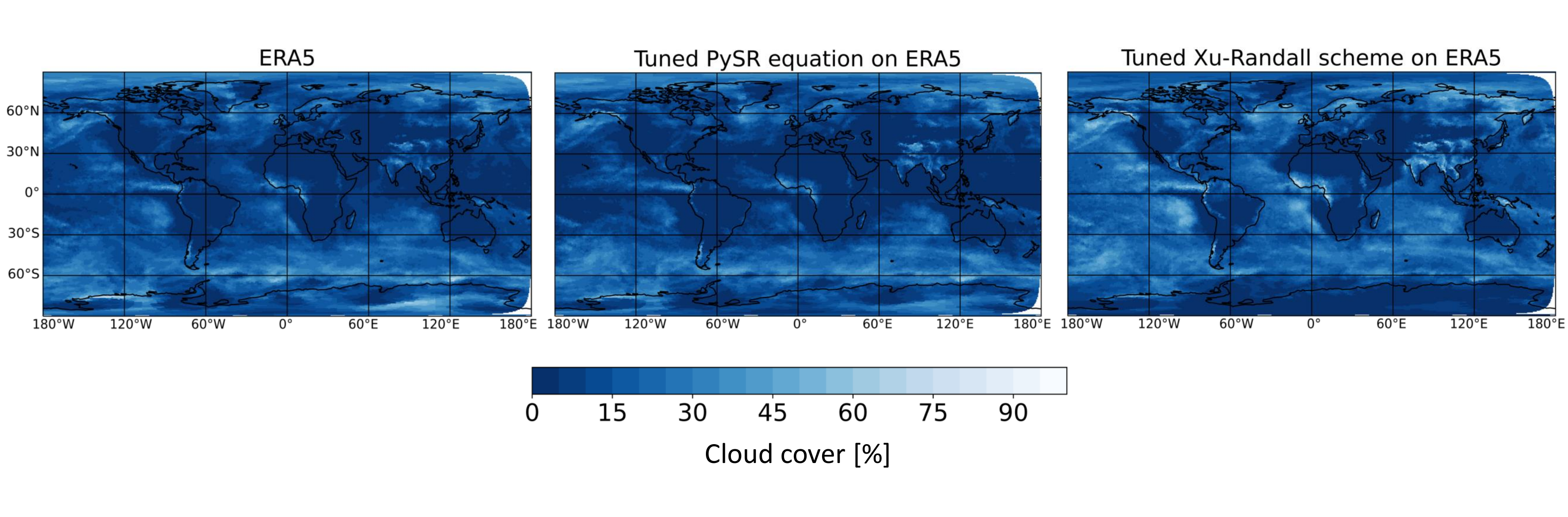}
    \caption{Averaged cloud cover from 11 Aug to 20 Aug, 2016 at an altitude of $\approx\,$1500 m. The data source is the remapped three-hourly ERA5 reanalysis data.}
    \label{fig:era5_1500m}
\end{figure}